\documentclass[11pt]{tibop-article}
\TIBauthor[Riehl et al.]{Kevin Riehl\affOne\orcidlink{0000-0003-4620-8379}\and
            Anastasios Kouvelas\affOne\orcidlink{0000-0003-4571-2530}\lastand 
            Michail A. Makridis\affOne\orcidlink{0000-0001-7462-4674}
          }
\TIBaffiliations{\affOne Traffic Engineering Group, Institute for Transport Planning and Systems, ETH Zurich, Stefano-Franscini-Platz 5, Zurich, 8093, Switzerland \medskip\\
	*Correspondence: Kevin Riehl, kriehl@ethz.ch
}

\TIBtitle{sumoITScontrol: Traffic Controller Collection for SUMO Traffic Simulations}
\TIBbundlename[SUMO User Conference 2026]{SUMO User Conference 2026}
\TIBconferencesession{Conference paper}%ignored for journal articles
\TIBabstract{
Reliable benchmarking is essential for progress in intelligent traffic control research. While microscopic traffic simulators such as SUMO enable detailed modelling of individual vehicle interactions, many published control studies still rely on single-run evaluations and project-specific baseline implementations, limiting reproducibility and comparability.
This paper presents \texttt{sumoITScontrol}, an open-source and extensible Python framework providing a curated collection of widely used traffic controllers implemented for SUMO via the TraCI interface. The framework includes established methods for both urban and freeway traffic management, such as Max Pressure signal control, SCOOT/SCATS-inspired adaptive strategies, and ramp metering algorithms including ALINEA, HERO-inspired, and METALINE.
Beyond providing implementations, the paper emphasises methodological best-practices for controller evaluation in stochastic microscopic environments. Through systematic calibration and replicated simulation experiments, we demonstrate the substantial impact of stochastic variability on performance metrics and highlight the necessity of variance-aware reporting and statistical hypothesis testing. 
By combining standardised controller implementations with reproducibility-oriented evaluation guidelines, \texttt{sumoITScontrol} aims to improve methodological transparency, enable fair benchmarking of novel approaches, and strengthen experimental standards within the SUMO and intelligent transportation systems research communities. \\
Source Code on project's GitHub:  \url{https://github.com/DerKevinRiehl/sumoITScontrol/}.
}
\TIBkeywords{intelligent transportation systems, traffic control, control theory, ITS, benchmarking}
\usepackage{xfrac}
\usepackage{booktabs}
\usepackage{amssymb}
\usepackage{pifont}
\usepackage{placeins}
\usepackage{microtype}
\begin{document}
\maketitle

% #######################################################################################
% #######################################################################################

\section{Introduction}

Efficient traffic management remains a central challenge in modern transportation systems. 
Growing travel demand, limited infrastructure capacity, and increasing societal expectations regarding sustainability and safety require control strategies that dynamically adapt to evolving traffic conditions. 
Intelligent Transportation Systems (ITS)~\cite{dimitrakopoulos2010intelligent} address these challenges by leveraging real-time sensor data~\cite{guerrero2018sensor} and algorithmic decision-making to dynamically adapt traffic control measures such as signal timing, perimeter control (urban context), and ramp metering rates and variable speed limits (freeway context).

Over the past decades, a wide range of traffic control strategies has been developed on the basis of macroscopic traffic flow theory. 
Classical approaches such as the store-and-forward model~\cite{GazisPotts1963_StoreAndForward} for urban networks and the METANET~\cite{Papageorgiou1990_METANET,hegyi2005model} and cell transmission model~\cite{daganzo1994cell} for freeway systems have provided the foundation for widely adopted controllers. 
Controllers such as Max-Pressure~\cite{varaiya2013max}, SCOOT~\cite{hunt1981scoot}, SCATS~\cite{lowrie1990scats}, ALINEA~\cite{papageorgiou1991alinea}, HERO~\cite{papamichail2010heuristic}, and METALINE~\cite{papageorgiou1990modelling} have demonstrated strong theoretical and practical performance under aggregated traffic representations, and have become established reference methods in the literature. 
However, macroscopic models inherently rely on simplifying assumptions regarding traffic homogeneity, flow continuity, and equilibrium dynamics. 
They abstract away individual vehicle behaviour and microscopic interactions, which can significantly influence congestion formation, shock-wave propagation, spill-back, and intersection blocking. 

In parallel, microscopic traffic simulation has become an increasingly important tool for analysing and evaluating control strategies. 
The open-source simulator SUMO (Simulation of Urban MObility)~\cite{8569938} enables detailed modelling of individual vehicle behaviour, including car-following dynamics, lane-changing decisions, and stochastic departure processes. 
Such microscopic representations allow researchers to assess controller performance under realistic traffic interactions, heterogeneous driver behaviour, and non-equilibrium dynamics that are difficult to capture in aggregated models.

As a consequence, a growing body of research proposes and evaluates new traffic control strategies directly within microscopic simulation environments~\cite{yang1996microscopic,barcelo2005microscopic}. 
These include analytical extensions of classical control methods as well as learning-based approaches such as reinforcement learning. 
While many studies report promising improvements, systematic benchmarking against established baseline controllers is often limited~\cite{riehl2025revisiting}. 
In numerous cases, baseline implementations are project-specific, insufficiently documented, or not publicly available, making replication and comparison difficult.
Implementing classical controllers within SUMO is non-trivial: it requires careful translation of theoretical formulations into discrete-time, detector-based control logic using interfaces such as TraCI. 
Differences in detector placement, aggregation intervals, and network modelling can further complicate fair comparisons~\cite{riehl2025eq}.

A further challenge arises from the inherently stochastic nature of microscopic simulation. 
Variability in departure times, route choices, car-following behaviour, and lane-changing decisions introduces non-negligible variance in performance metrics such as travel time, delay, and throughput. 
Small reported improvements may therefore lie within the range of stochastic fluctuations rather than reflecting statistically significant performance gains.
Nevertheless, many simulation-based studies rely on single-run evaluations or fail to report variance measures, limiting the robustness and interpretability of their conclusions.

These challenges highlight the need for a standardised, reproducible benchmarking framework for traffic control research in SUMO~\cite{riehl2025revisiting}. 
Such a framework should provide: (i) transparent and well-documented implementations of established controllers, (ii) modular interfaces enabling consistent experimental design, and (iii) methodological guidance for variance-aware calibration and statistical evaluation.

This paper introduces \texttt{sumoITScontrol}, an open-source and extensible Python package implementing a curated collection of widely used traffic controllers for SUMO via the TraCI interface. 
The framework covers both urban and freeway control applications. For signalised intersections, it includes controllers such as Max Pressure and adaptive coordinated strategies inspired by SCOOT and SCATS principles. 
For freeway traffic management, it provides established ramp metering algorithms including ALINEA, METALINE, and HERO-inspired coordinated controllers. 
All controllers are implemented in a modular and configurable manner to facilitate consistent benchmarking across different network scenarios.

Beyond providing software implementations, this work emphasises methodological best-practices for simulation-based controller evaluation. 
Through structured calibration procedures, replicated simulation experiments, and formal statistical hypothesis testing, we demonstrate the substantial impact of stochastic variability on controller performance assessment. 
The presented case studies illustrate how parameter optimisation and controller comparison can be conducted in a variance-aware and statistically sound manner.

The primary contribution of this paper is therefore twofold: first, the provision of a transparent and extensible controller collection for SUMO; and second, the establishment of a reproducibility-oriented evaluation paradigm for microscopic traffic control research. 
By lowering the barrier to rigorous benchmarking and encouraging statistically robust experimentation, \texttt{sumoITScontrol} aims to strengthen methodological standards and improve comparability across studies within the SUMO and ITS research communities.

The remainder of this work is organised as follows.
Section~2 outlines the controller implementation.
Section~3 provides guidelines on simulation-design and sensor placement.
Section~4 elaborates on best-practices for stochastic controller calibration and evaluation.
Section~5 presents the results of simulation experiments with the two provided case studies.
Finally, Section~6 concludes the paper.

\section{Controller Implementation} \label{control}

\subsection{Ramp Metering}

One of the most widely implemented highway traffic control strategies is ramp metering, which regulates vehicle entry rates at highway on-ramps (with traffic lights) to ensure an uncongested traffic on the main-road (optimizing for total travel time, or total travel distance). 
First introduced in the 1960s on Chicago’s \textit{Eisenhower Expressway}, ramp metering has since become a core component of freeway traffic management in major U.S. cities, including Los Angeles, Minneapolis, and Chicago~\cite{papageorgiou2002freeway}. 

\subsubsection{Problem Statement \& Notation}

\begin{figure}[!h]
    \centering
    \includegraphics[width=0.5\linewidth]{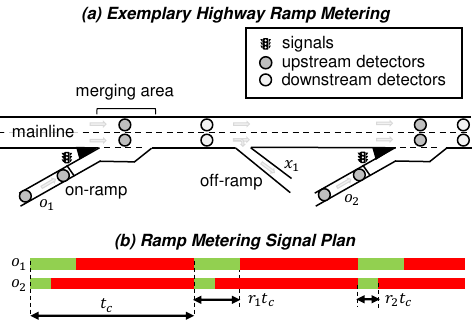}
    \caption{Ramp Metering Problem Statement.~\cite{riehl2025eq}}
    \label{fig:ramp_metering}
\end{figure}

The highway consists of mainline lanes, and a set of on-ramps $o_i \in \mathcal{O}$ and off-ramps $x_i \in \mathcal{X}$.
Each metered on-ramp is equipped with a traffic light, that follows a signal cycle of duration $t_c$, from which the metering rate $r_i \in [0, 1 ]$ represents the share of green duration of $t_c$.
When the traffic lights turn green, vehicles waiting on the on-ramp can continue to drive onto the highway, and merge with the mainline traffic in the merging area, as shown in Figure~\ref{fig:ramp_metering}.

Depending on traffic volume and driving behaviour, this merging process can cause congestion, and back-propagate upstream, as the merging area represents a bottleneck for the traffic.
Other sources of congestion are blocked off-ramps or lane changes before those, and back-propagation of stop-and-go shock-waves (known as phantom traffic jam).
There are sensors $j$ (loop-detectors) that can measure traffic flow characteristics on the mainline, namely flow $f_t$, average vehicle speed $v_t$, and occupancy $c_t$, and physical queue lengths at on-ramps $l_t$, at regular, discrete time intervals $t$.
Some ramp-metering algorithms use upstream detectors, that are located in the merging areas, while others use downstream detectors, that are located downstream, after the merging area.
The control problem of ramp-metering is to adjust ramp metering rates $r_i$ dynamically according to the traffic conditions on the highway, to achieve certain goals, such as transportation efficiency (often measured in total travel time).
Often, it can make sense to bound metering rates to physically plausible ranges $r_i \in [r_{min}, r_{max}]$, e.g. $r_{min}=0$\%, and $r_{max}=100$\%. 
Arguably, lower bounds can also be set above 0\% to avoid massive queues or spill-back into urban road networks that feed the highways. 

\begin{table}[!h]
    \centering
    \caption{\textbf{Notation for Ramp Metering.}}
    \label{tab:placeholder}
    \small
    \begin{tabular}{ll}
        \toprule
        \textbf{Symbol} & \textbf{Parameter} \\
        \midrule
        \multicolumn{2}{l}{\textbf{Indices}} \\
        $i$ & ramp index \\
        $j$ & sensor index \\
        $t$ & discrete time index \\ \\
        \multicolumn{2}{l}{\textbf{Descriptors}} \\
        $x_i$ & off-ramp, part of subset $\mathcal{X}$ \\
        $o_i$ & on-ramp, part of subset $\mathcal{O}$\\
        $\mathcal{X}$ & set of all off-ramps \\
        $\mathcal{O}$ & set of all on-ramps \\ \\
        \multicolumn{2}{l}{\textbf{Sensing Variables}} \\
        $f_{t,i}$ & measured vehicle flow at ramp $i$ at time $t$\\
        $v_{t,i}$ & measured average vehicle speed at ramp $i$ at time $t$\\
        $c_{t,i}$ & measured vehicle density at ramp $i$ at time $t$ \\
        $l_{t,i}$ & measured queue length ramp $i$ at time $t$ \\ \\
        \multicolumn{2}{l}{\textbf{Actuation Variables}} \\
        $r_i$ & metering rate of ramp $i$, share of $t_c$ where signal is green, bounded by $r_i \in [r_{min}, r_{max}]$\\
        $r_{min}$ & minimum bound for $r_i$, typically $\geq0$\% \\
        $r_{max}$ & maximum bound for $r_i$, typically $\leq100$\% \\
        $t_c$ & signal cycle duration \\ \\
        \multicolumn{2}{l}{\textbf{Control Parameters}} \\
        $c^{\star}$ & desired target occupancy \\
        $K_P$ & regulator gain (tuning parameter) \\
        $K_I$ & integrator gain (tuning parameter) \\
        $l^{act}$ & actuation threshold, critical queue length \\
        $n$ & number of coordinated ramps \\
        $N_{t,i}$ & estimated number of vehicles in queue \\
        $N_{i}^{max}$ & desired maximum queue set-point \\
        $\hat{q}_{t,i}^{in}$ & estimated arrival flow at ramp (HERO) \\
        $q_i^{sat}$ & saturation discharge flow (HERO) \\
        $q_{t,i}^{ctrl}$ & desired discharge flow (HERO) \\
        $\alpha$ & anticipation factor (HERO) \\
        \bottomrule
    \end{tabular}
\end{table}

\subsubsection{ALINEA}

A widely adopted local feedback ramp-metering strategy is \textbf{ALINEA}, which stands for \emph{Asservissement Linéaire d’Entrée Autoroutière} (French for “linear feedback control of freeway entrance”).
ALINEA was originally proposed by \textit{Markos Papageorgiou}~\cite{papageorgiou1991alinea} in the early 1990s and has since become one of the most influential and practically implemented ramp metering algorithms in freeway traffic management systems.
ALINEA is a simple integral feedback controller that regulates the ramp metering rate based on measured mainline occupancy downstream of the merge area. 
The key idea is to maintain the measured occupancy $c_t$ close to a desired target occupancy $c^{\star}$, which can corresponds to the maximum flow (i.e., capacity) of the freeway. 
By stabilizing traffic density near this target working point, ALINEA aims to prevent congestion formation, maximise throughput, and also enhance traffic safety through density homogenisation. The classical ALINEA controller uses a simple P-feedback law given by:
\begin{equation}
    r_t = r_{t-1} + K_p (c^{\star} - c_t)
\end{equation}
where $K_P$ is a regulator gain (tuning parameter).

While the original \textit{ALINEA} leveraged downstream occupancy sensors (40 - 100m downstream), other local strategies include the usage of upstream sensors \textit{(UP-ALINEA}~\cite{papageorgiou2008misapplication,smaragdis2003series}) or both upstream and downstream sensors (\textit{MALINEA}~\cite{oh2000study}), flow measurements instead off occupancy (\textit{FL-ALINEA}~\cite{smaragdis2003series}), and a PI-control law (\textit{PI-ALINEA}~\cite{demiral2011application,abuamer2018comparative}), which is offered by our implementation as well:
\begin{equation}
    r_t = r_{t-1} + K_p (c^{\star} - c_t) + K_I (c_t - c_{t-1})
\end{equation}
where $K_I$ is an integrator gain (tuning parameter).

\subsubsection{METALINE}

While ALINEA is a local feedback controller that regulates each on-ramp independently, METALINE~\cite{papageorgiou1990modelling} extends this concept to a coordinated, multi-ramp control framework.
METALINE can be interpreted as a multivariable generalisation of ALINEA, in which several metered on-ramps are jointly controlled in order to account for spatial interactions along the mainline.

Consider a set of $n$ metered on-ramps $o_i \in \mathcal{O}$.
Let the vectors of metering rates, occupancies, and target occupancies be defined as
\begin{equation}
    \mathbf{r}_t = 
    \begin{bmatrix}
        r_{t,1} & \dots & r_{t,n}
    \end{bmatrix}^\top, 
    \qquad
    \mathbf{c}_t = 
    \begin{bmatrix}
        c_{t,1} & \dots & c_{t,n}
    \end{bmatrix}^\top,
    \qquad
    \mathbf{c}^{\star} =
    \begin{bmatrix}
        c^{\star}_1 & \dots & c^{\star}_n
    \end{bmatrix}^\top.
\end{equation}

In contrast to independent local feedback, METALINE introduces coupling between ramps through gain matrices $\mathbf{K}_P \in \mathbf{R}^{n \times n}$ and $\mathbf{K}_I \in \mathbf{R}^{n \times n}$. 
The coordinated PI-type feedback law is given by:
\begin{equation}
    \mathbf{r}_t 
    =
    \mathbf{r}_{t-1}
    +
    \mathbf{K}_P
    \left(
        \mathbf{c}^{\star} - \mathbf{c}_t
    \right)
    +
    \mathbf{K}_I
    \left(
        \mathbf{c}_t - \mathbf{c}_{t-1}
    \right).
\end{equation}

The off-diagonal elements of $\mathbf{K}_P$ and $\mathbf{K}_I$ allow upstream ramps to react to downstream congestion levels and vice versa, thereby capturing the spatial propagation of traffic disturbances. 
This coordinated regulation can improve global traffic performance compared to purely local control, particularly in corridors with closely spaced ramps.
As in ALINEA, the metering rates are bounded to physically admissible limits,
\begin{equation}
    r_{t,i} \in [r_{min}, r_{max}], 
    \qquad \forall i \in \{1, \dots, n\},
\end{equation}
in order to avoid excessive queue growth or spill-back into the surrounding urban network.
In the special case where $\mathbf{K}_P$ and $\mathbf{K}_I$ are diagonal matrices, METALINE reduces to a set of independent PI-ALINEA controllers.

\subsubsection{HERO}

HERO (Heuristic Ramp metering co-Ordination)~\cite{papamichail2010heuristic} is a hierarchical coordination strategy that augments local ALINEA controllers with an additional queue management layer.
The primary objective of HERO is to prevent excessive queue formation at individual on-ramps by redistributing metering restrictions upstream.
Under normal operating conditions, each ramp $o_i \in \mathcal{O}$ is regulated by its local ALINEA controller.
However, when the queue length $l_{t,i}$ of a ramp exceeds a predefined activation threshold $l^{act}$, a coordinated control mode is triggered.

\textit{Master--slave structure:} 
Once activated, the ramp with excessive queue length is designated as the \emph{master} ramp.
Upstream ramps may be progressively recruited as \emph{slave} ramps.
While the master ramp continues to operate under ALINEA, slave ramps temporarily override their local ALINEA metering rates and apply a queue-based control law in order to restrict inflow to the master region.

\textit{Queue-based control of slave ramps:} For a slave ramp $o_i$, the queue length $l_{t,i}$ (in metres) is converted into an approximate number of vehicles using an average vehicle spacing parameter.
Let $N_{t,i}$ denote the estimated number of vehicles in the queue and $N^{max}_i$ a desired maximum queue set-point.
A simplified deadbeat-style control objective aims to drive the queue towards $N^{max}_i$ within one signal cycle of duration $t_c$.
Denoting by $\hat{q}_{t,i}^{in}$ the estimated arrival flow to the ramp and by $q_i^{sat}$ the saturation discharge flow, the desired discharge flow is computed as
\begin{equation}
    q_{t,i}^{ctrl}
    =
    \frac{
        N^{max}_i - N_{t,i} + \alpha \hat{q}_{t,i}^{in}
    }{t_c},
\end{equation}
where $\alpha$ is an anticipation factor accounting for expected arrivals.
The corresponding metering rate is then obtained as
\begin{equation}
    r_{t,i}
    =
    \frac{
        q_{t,i}^{ctrl}
    }{
        q_i^{sat}
    },
\end{equation}

subject to the bounds

\begin{equation}
    r_{t,i} \in [r_{min}, r_{max}].
\end{equation}

\textit{Dissolution condition:}
The coordinated mode is terminated once the master queue length falls below a release threshold $l^{rel}$, with $l^{rel} < l^{act}$ in order to avoid chattering.
All ramps then revert to independent ALINEA operation.
Through this hierarchical master--slave mechanism, HERO preserves the throughput optimisation properties of ALINEA while explicitly mitigating queue spill-back and improving robustness in heavily congested scenarios.

\subsubsection{Exclusion of Other Controllers}

In addition to local and coordinated feedback strategies such as ALINEA, METALINE, and HERO, large-scale adaptive ramp metering systems have been implemented in practice, most notably the SWARM~\cite{Monsere2008UsingAI} and ZONE~\cite{Stephanedes1994ImplementationOO} algorithms deployed in the Minneapolis-Saint Paul freeway network. 
These approaches operate on a corridor or network-wide level and rely on extensive detector infrastructure, centralised optimisation routines, and complex calibration procedures based on archived ITS data. 
Their control logic typically involves dynamic estimation of traffic states over extended freeway sections and the computation of coordinated metering rates for predefined zones.

In the present work, SWARM and ZONE-based strategies are excluded for two main reasons. 
First, their implementation requires a substantially higher level of system-wide modelling, calibration effort, and data availability than is assumed in our framework, which focuses on modular, ramp-level and corridor-level feedback control using locally available measurements $f_{t,i}$, $c_{t,i}$, and $l_{t,i}$. 
Secondly, the objective of this study is to analyse and compare structurally transparent feedback controllers with clearly interpretable parameters (e.g.\ $K_P$, $K_I$, $c^{\star}$), rather than large-scale adaptive systems whose performance strongly depends on network-specific tuning and historical data processing. 
By restricting the scope to ALINEA-type and HERO-type controllers, the analysis remains reproducible, methodologically consistent, and transferable to different motorway configurations without requiring extensive re-calibration.

\subsection{Signalised Intersection Management}

Signalised intersection management is one of the oldest, most widespread forms of urban traffic control and central to both efficiency and safety in cities, dating back to the installation of the first electric traffic signal in Cleveland in 1914, and later standardised with timed control systems in the mid-20th century. 
As cities grew denser and vehicle ownership increased, intersections emerged as critical bottlenecks in the urban transport network, where queue build-up and conflict between movements had to be carefully managed. 
Traditional fixed-time signal plans -- developed from early engineering heuristics -- gradually evolved toward actuated and adaptive systems, capable of responding to real-time traffic conditions. 
Nowadays, in many large metropolitan areas, several thousands of intersections are signalised, and almost all major arterial corridors are controlled by coordinated traffic signals~\cite{roess2004traffic,garber2009traffic}. 

\begin{figure}[!h]
    \centering
    \includegraphics[width=\linewidth]{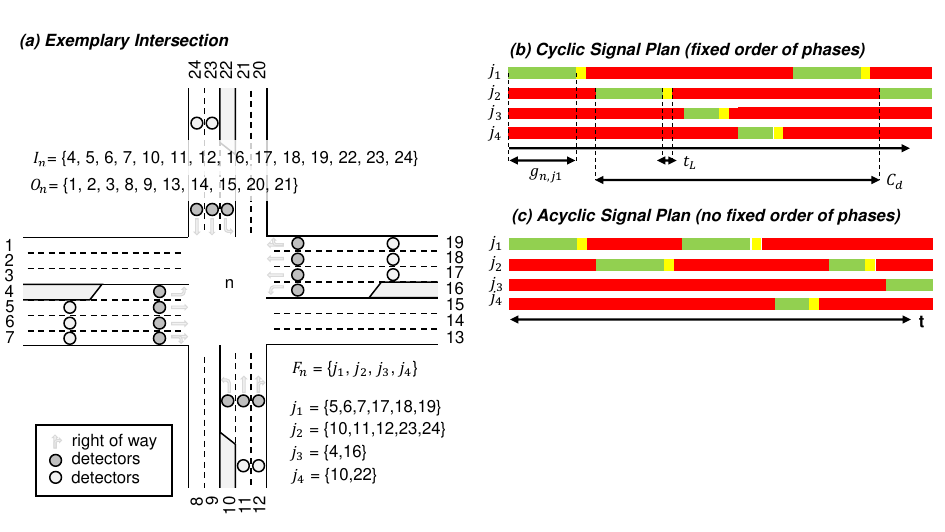}
    \caption{Signalised Intersection Management Problem Statement.~\cite{riehl2025green}}
    \label{fig:intersection_management}
\end{figure}

\subsubsection{Problem Statement \& Notation}

The road network is represented as a directed graph with nodes $n \in N$, that are intersections, and links $z\in Z$, that connected these intersections.
The set of incoming links ($I_n$) and outgoing links ($O_n$) are defined for each intersection $n$.
A set of non-conflicting links (approaches) $z_j$, that can simultaneously receive right-of-way without collision, forms a movement phase $j$.
The signal control plan of intersection $n$ assigns green durations $g_{n,j}(k_n)$ for each cycle $k_n$, based on a fixed set of possible movement phases $j \in F_n$. 

Each green-time follows minimum and maximum permissible green time constraints:
\begin{equation}
    g_{n,j,\min} \leq g_{n,j}(k_n) \leq g_{n,j,\max} \; \; \forall j \in F_n
\end{equation}

The transition from one to the next phase requires a safety-critical period of yellow and red signals.
Frequent switches can support prioritizing right of way for long queues, but come at the cost of lost transition times $t_L$. 

The state of each link $z$ can be described by pressure $p_z(k_n)$, which could be a function of the number of vehicles waiting in the queue to be served.
The pressure of a movement phase $j$ can therefore be described as: 
\begin{equation}
    p_j(k_n) = \sum_{z \in z_j} p_z(k_n)
\end{equation}

The controller's task can be summarised as to define a signal control plan with $g_{n,j}(k_n)$ given $p_j(k_n)$.

Figure~\ref{fig:intersection_management} showcases cyclic and acyclic signal plans at an exemplary intersection.
The order of phases $j_1, j_2, j_3, j_4$ is recurring in the cyclic plan, while right of way is dynamically assigned to any phase in the acyclic plan.

\begin{table}[!h]
    \centering
    \caption{\textbf{Notation for Intersection Management.}}
    \label{tab:placeholder2}
    \small
    \begin{tabular}{ll}
        \toprule
        \textbf{Symbol} & \textbf{Parameter} \\
        \midrule
        \multicolumn{2}{l}{\textbf{Indices}} \\
        $n$ & specific intersection, part of set $n \in N$\\
        $z$ & specific link, part of set $z \in Z$\\
        $j$ & movement phase, part of set $F_n$ \\ \\

        \multicolumn{2}{l}{\textbf{Descriptors}} \\
        $N$ & set of all nodes (intersections) \\
        $Z$ & set of all edges (links connecting the intersections) \\
        $\mathcal{N}_g$ & coordinated group of intersections \\
        $I_n$ & set of incoming links \\
        $O_n$ & set of outgoing links \\
        $F_n$ & set of possible movement phases \\
        $k_n$ & control cycle \\
        $z_j$ & set of links served in movement phase $j$ \\
        $j(t)$ & currently active phase at time $t$ \\ \\
        
        \multicolumn{2}{l}{\textbf{Sensing Variables}} \\
        $p_{z}(k_n)$ & pressure (e.g. number of vehicles waiting in a link z's queue to be served) \\
        $p_{j}(k_n)$ & pressure (e.g. number of vehicles waiting in a phase j's queue to be served) \\
        $\bar{p}_j(k_n)$ & average phase pressure within one cycle \\
        $d_z(k_n)$ & degree of saturation of lane $z$ \\
        $d_{\max}(k_n)$ & maximum degree of saturation in $\mathcal{N}_g$ \\ \\
        
        \multicolumn{2}{l}{\textbf{Actuation Variables}} \\
        $g_{n,j}(k_n)$ & green duration for cycle $k_n$, bounded by $g_{n,j}(k_n) \in [g_{n,j,min} , g_{n,j,max}]$ \\
        $\theta_n(k_n)$ & signal offset of intersection $n$ \\
        $t_c$ & signal cycle duration \\ \\

        \multicolumn{2}{l}{\textbf{Timing Parameters}} \\
        $g_{n,j,min}$ & minimum green phase duration \\
        $g_{n,j,max}$ & maximum green phase duration \\
        $t_L$ & (lost) transition time between movement phases (e.g. yellow time) \\
        $t_{\text{eff}}$ & effective green time per cycle \\
        $t_{\text{eff},n}$ & effective green time at intersection $n$ \\
        $t_c(k_n)$ & cycle duration at control step $k_n$ \\
        $t_{c,\min}$ & minimum cycle duration \\
        $t_{c,\max}$ & maximum cycle duration \\
        $t_a$ & adaptive waiting interval (Max-Pressure Flex) \\
        $H$ & number of pressure measurements per cycle \\ \\
        
        \multicolumn{2}{l}{\textbf{Scoot/Scats Parameters}} \\
        $\alpha_c$ & cycle length adaptation gain \\
        $\alpha_g$ & green split adaptation gain \\
        $\alpha_o$ & offset adaptation gain \\
        $d_{\text{upper}}$ & upper saturation threshold \\
        $d_{\text{lower}}$ & lower saturation threshold \\
        $\tau_{n \rightarrow m}$ & estimated travel time between intersections \\
        $L_{n \rightarrow m}$ & connection length between intersections \\
        $v_{\text{limit}}$ & assumed progression speed \\
        
        \bottomrule
    \end{tabular}
\end{table}

%\FloatBarrier

\subsubsection{Max-Pressure (Introduction) }

Max-Pressure is a controller that aims to release the "pressure" (analogy to vehicles accumulated in queues) of movement phases at intersections as fast as possible.
\textit{Tassiulas et al.}~\cite{tassiulas1990stability} first presented the Max-Pressure algorithm for routing and scheduling packet transmission in wireless networks.
\textit{Variaya et al.}~\cite{varaiya2013max} introduced the Max-Pressure controller into the domain of signal control, and defined pressure as queue lengths in number of vehicles.
\textit{Kouvelas et al.}\cite{kouvelas2014maximum} extended the Max Pressure concept, and constrained the problem by several aspects, such as a strict cyclic phase policy and cycle duration, and inclusion of turning-ratio knowledge. 
Moreover, normalised pressure (queue lengths divided by storage capacity of the link) were considered to address the threat of spill backs to neighbouring intersections.
Further more, works explored the use of input travel times as link pressures and consider bounded capacity links~\cite{mercader2020max}, enforced a fixed phase order, and proposed to reduced the update frequency of green phase durations~\cite{anderson2018stability}, or included boundary conditions for the number of switches within one cycle~\cite{pumir2015stability}.
In addition to that, a broad variety of economic, auction based signal controllers have been proposed, that implement non-cyclic phase policies, describing the phase pressures as bids in an auction~\cite{iliopoulou2022survey}, similar to ~\cite{varaiya2013max}.
Figure~\ref{fig:intersection_management} showcases cyclic and acyclic signal plans at an exemplary intersection.
The order of phases $j_1, j_2, j_3, j_4$ is recurring in the cyclic plan, while right of way is dynamically assigned to any phase in the acyclic plan.

The different variants of Max-Pressure algorithms vary in:
\begin{itemize}
    \item their definition of pressure, 
    \item their implementation of strict cyclic phase policies, 
    \item their implementation of cycle durations, and 
    \item additional boundary conditions.
\end{itemize}

% The Max-Pressure control algorithm can be decomposed into three decision making processes~\cite{kouvelas2014maximum}:
% \begin{itemize}
%     \item offset optimisation between cycles of different intersections to enable green waves,
%     \item cycle duration $C_d$ optimisation, and 
%     \item cycle split $\Delta_{n1, n2}$ optimisation (share of green times assigned to each phase).
% \end{itemize}

Within \texttt{sumoITScontrol} we implement two versions of Max-Pressure, one with a strict fixed-order phase cycle, and one with a flexible phase assignment.

\subsubsection{Max-Pressure (Fixed) }

The fixed-order Max-Pressure controller implements a cyclic phase policy with a fixed cycle duration $t_c$ and a predefined order of movement phases $j \in F_n$. 
Green times are reallocated at the beginning of each cycle proportionally to the measured phase pressures $p_j(k_n)$, while the phase sequence itself remains unchanged.
The fixed-order Max-Pressure variant therefore exhibits the following properties:
\begin{itemize}
    \item Strict cyclic phase policy.
    \item Fixed cycle duration $t_c$.
    \item Pressure-proportional cycle split optimisation.
    \item Deterministic inter-green time $t_L$ between phases.
    \item No dynamic reordering of phases within a cycle.
\end{itemize}
This formulation corresponds to a constrained variant of Max-Pressure control as discussed in~\cite{kouvelas2014maximum}, where adaptivity is restricted to the allocation of green splits while phase order and cycle duration remain fixed.

\textit{Pressure Measurement:}
During each cycle $k_n$, pressures $p_j(k_n)$ are sampled periodically and stored over the cycle horizon. 
At the end of the cycle, the average pressure for each phase is computed as:

\begin{equation}
\bar{p}_j(k_n) = \frac{1}{H} \sum_{h=1}^{H} p_j(k_n, h),
\end{equation}

where $H$ denotes the number of measurement intervals within one cycle.
In the current implementation, $p_j(k_n)$ corresponds to the number of vehicles waiting in the queues of all links belonging to phase $j$.

\textit{Cycle Split Optimisation:}
Given $n = |F_n|$ phases and a fixed transition time $t_L$ per phase change, the total lost time per cycle is:

\begin{equation}
t_{\text{loss}} = n \cdot t_L.
\end{equation}

The effective green time available within one cycle is therefore:

\begin{equation}
t_{\text{eff}} = \max \left(0, t_c - t_{\text{loss}} \right).
\end{equation}

Initial green splits are assigned proportionally to the measured pressures:

\begin{equation}
g_{n,j}^{(0)}(k_n) =
\begin{cases}
\dfrac{\bar{p}_j(k_n)}{\sum_{i \in F_n} \bar{p}_i(k_n)} \cdot t_{\text{eff}}, & \text{if } \sum_i \bar{p}_i(k_n) > 0, \\
\dfrac{t_{\text{eff}}}{n}, & \text{otherwise}.
\end{cases}
\end{equation}

Minimum and maximum green constraints are enforced:

\begin{equation}
g_{n,j}(k_n) \in [g_{n,j,\min}, g_{n,j,\max}].
\end{equation}

If constraint enforcement causes the total allocated green time to differ from $t_{\text{eff}}$, remaining time is redistributed iteratively among phases that can be increased or decreased, until the sum matches $t_{\text{eff}}$. 
Finally, green times are rounded to integer seconds:

\begin{equation}
g_{n,j}(k_n) \in \mathbb{Z}_{\ge 0}, \quad \sum_{j \in F_n} g_{n,j}(k_n) = \lfloor t_{\text{eff}} \rceil.
\end{equation}

\textit{Signalling:}
The controller signals sequentially through the fixed phase order $j_1, j_2, \dots, j_n$ with wrap-around after the final phase:

\begin{equation}
\text{Green}(j_1) \rightarrow \text{Transition} \rightarrow \text{Green}(j_2) \rightarrow \cdots \rightarrow  \text{Green}(j_n) \rightarrow  \text{Green}(j_1),
\end{equation}

A full cycle is completed once all phases in $F_n$ have been served exactly once. At this point, the controller returns to the \textit{Idle} state, recomputes $g_{n,j}(k_{n+1})$, and initiates the next cycle.

\subsubsection{Max-Pressure (Flexible) }

The flexible Max-Pressure controller removes the strict cyclic phase order and fixed cycle duration assumed in the previous variant. 
Instead, right-of-way is dynamically assigned to the movement phase with the highest pressure, subject to minimum and maximum green constraints. 
The controller therefore implements an acyclic, event-driven phase policy.
This formulation more closely resembles the classical back-pressure routing logic introduced in~\cite{tassiulas1990stability,varaiya2013max}, where service is allocated to the queue with the highest instantaneous pressure. 
However, unlike the theoretical continuous-time formulation, the implementation explicitly incorporates safety inter-green times $t_L$, discrete measurement intervals, and bounded green durations, rendering it directly applicable within microscopic simulation environments such as SUMO.
Compared to the fixed-order variant, the flexible Max-Pressure controller exhibits:
\begin{itemize}
    \item No predefined cycle duration $t_c$.
    \item No fixed phase sequence.
    \item Event-driven switching based on instantaneous pressure comparisons.
    \item Enforcement of minimum and maximum green constraints.
    \item Random tie-breaking among equally pressured phases.
\end{itemize}

\begin{figure}
    \centering
    \includegraphics[width=0.5\linewidth]{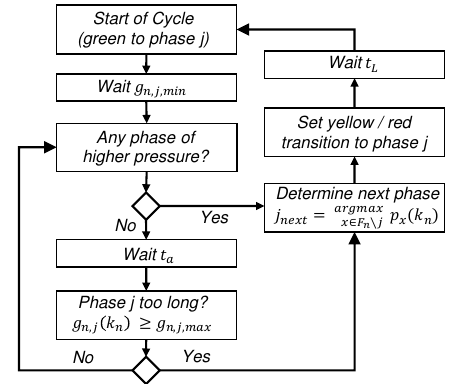}
    \caption{\textbf{Max-Pressure Algorithm as Finite State Machine.}}
    \label{fig:fsm_max}
\end{figure}

\textit{Pressure Measurement:}
At every measurement interval, the phase pressures $p_j(k_n)$ are observed. 
In the present implementation, pressure corresponds to the number of vehicles waiting on the incoming links associated with phase $j$.
Let the currently active green phase at time $t$ be $j(t)$. 
The corresponding pressure is denoted as $p_{j(t)}(k_n)$.

\textit{Decision Logic:}
The controller enforces a minimum green time $g_{n,j,\min}$ for each activated phase. 
After this minimum duration has elapsed, the controller evaluates whether another phase exhibits higher pressure:

\begin{equation}
\max_{i \in F_n} p_i(k_n) > p_{j(t)}(k_n). \label{equation}
\end{equation}

If this condition holds, the controller initiates a phase change towards one of the phases achieving the maximum pressure. 
In the case of multiple maximisers, one is selected randomly to avoid systematic bias.

If no competing phase exhibits higher pressure, the controller maintains the current phase for an additional adaptive interval $t_a$. 
This re-evaluation procedure continues until either:

\begin{itemize}
    \item a competing phase exhibits strictly higher pressure, or
    \item the maximum green duration $g_{n,j,\max}$ is reached.
\end{itemize}

Thus, green times are not pre-computed but emerge endogenously from repeated pressure comparisons.

\textit{Finite State Machine Representation:}
The flexible controller is implemented as a finite state machine (FSM) with five states, as shown in Figure~\ref{fig:fsm_max}:
\begin{itemize}
    \item \textbf{Start}: a new green phase has just been activated; the controller enforces the minimum green time $g_{n,j,\min}$.
    \item \textbf{Check\_Pressures}: the pressure of the current phase is compared with all competing phases.
    \item \textbf{Wait}: no competing phase has higher pressure; the controller waits for $t_a$ before re-evaluating.
    \item \textbf{Next\_Phase}: a new phase with maximal pressure is selected.
    \item \textbf{Transition}: safety-critical inter-green period of duration $t_L$.
\end{itemize}

The state transitions can be summarised as:

\begin{align}
\text{Start} &\rightarrow \text{Check\_Pressures}, \\
\text{Check\_Pressures} &\rightarrow 
\begin{cases}
\text{Next\_Phase}, & \text{if } \exists i: p_i > p_{j(t)} \\
\text{Wait}, & \text{otherwise}
\end{cases} \\
\text{Wait} &\rightarrow 
\begin{cases}
\text{Next\_Phase}, & \text{if } g_{n,j}(t) \geq g_{n,j,\max} \\
\text{Check\_Pressures}, & \text{otherwise}
\end{cases} \\
\text{Next\_Phase} &\rightarrow \text{Transition} \\
\text{Transition} &\rightarrow \text{Start}.
\end{align}

After the transition period $t_L$, the newly selected phase becomes active and the process repeats.

\subsubsection{SCOOT/SCATS}

\begin{figure}
    \centering
    \includegraphics[width=\linewidth]{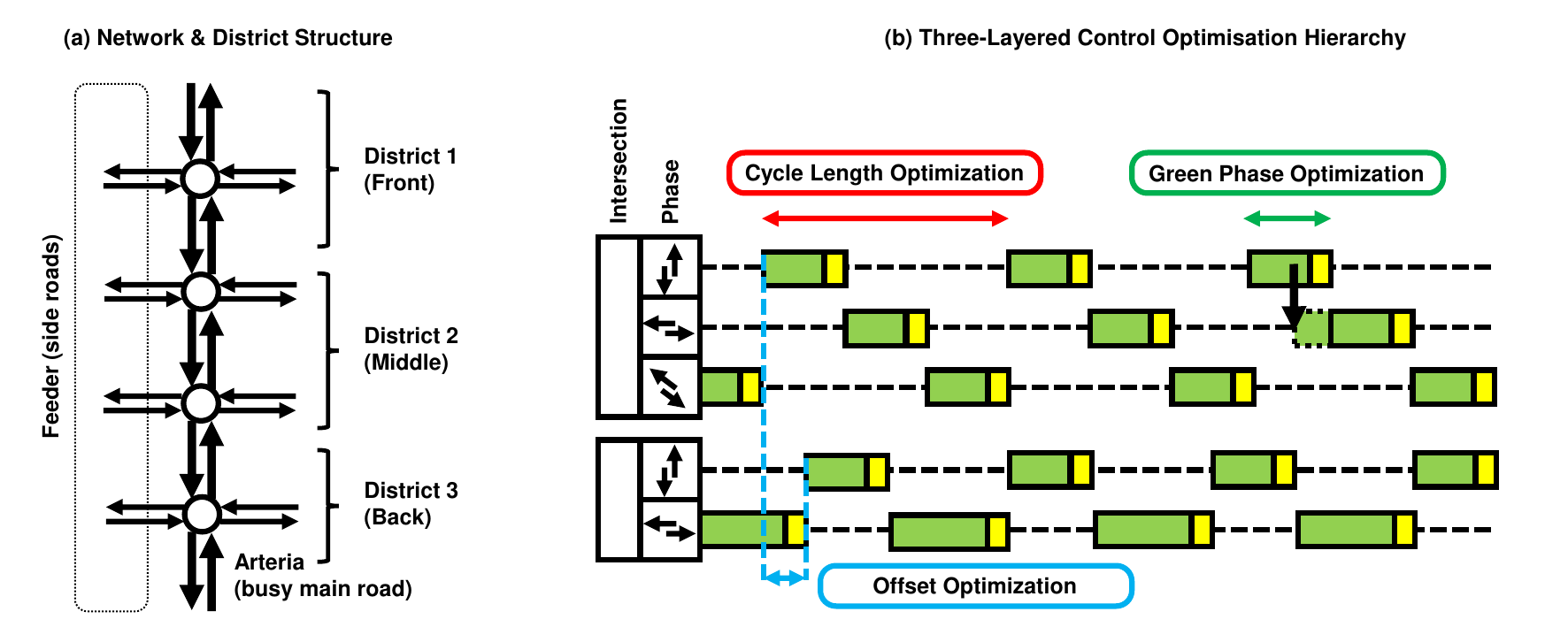}
    \caption{\textbf{Scoot/Scats Controller For Coordinated Signalised Intersection Management.}}
    \label{fig:scosca}
\end{figure}

In contrast to the previously described intersection-local Max-Pressure controllers, the implemented SCOOT/SCATS-inspired controller operates on a group of coordinated intersections and follows a hierarchical three-step optimisation procedure as outlined in Figure~\ref{fig:scosca}:

\begin{enumerate}
    \item Cycle length optimisation,
    \item Green split optimisation,
    \item Offset optimisation.
\end{enumerate}

This structure reflects the classical philosophy of adaptive urban traffic control systems such as SCOOT~\cite{hunt1981scoot} and SCATS~\cite{lowrie1990scats}, where coordination across corridors is achieved through gradual parameter adaptation rather than instantaneous phase switching.
The implemented SCOOT/SCATS variant exhibits:
\begin{itemize}
    \item Group-based control of multiple intersections.
    \item Shared adaptive cycle length.
    \item Local green split optimisation driven by lane saturation.
    \item Corridor-level offset optimisation based on estimated travel times.
    \item Gradual parameter adaptation rather than event-driven switching.
\end{itemize}
Unlike the Max-Pressure controllers, which allocate right-of-way reactively based on instantaneous pressure, the SCOOT/SCATS controller performs hierarchical optimisation at cycle boundaries, prioritising network-wide coordination and progression over short-term queue minimisation.

\textit{Network Setting:}
Let $\mathcal{N}_g \subseteq N$ denote a coordinated group of intersections.
All intersections in $\mathcal{N}_g$ share a common cycle length $t_c$, while green splits and offsets are defined individually:

\begin{itemize}
    \item $g_{n,j}(k_n)$: green split for phase $j$ at intersection $n$,
    \item $\theta_n(k_n)$: offset of intersection $n$ relative to the group reference,
    \item $t_c(k_n)$: common cycle length of the group.
\end{itemize}

Measurements consist of two lane-level components, (i) queue lengths, and (ii) degree of saturation$d_{z}(k_n)$ per lane $z$. 
In SCOOT and SCATS, the degree of saturation expresses how "busy" a lane, approach, or movement is relative to its usable capacity during the signal cycle. 
It is usually given as a ratio or percentage, where 1.0 (or 100\%) means demand is equal to effective capacity and the approach is operating at or very near saturation, with values above this indicating over-saturation and growing queues.
The optimisation is triggered at the end of each cycle, and outlined in the following.

% --------------------------------------------------
\textit{1. Cycle Length Optimisation:}
The first step adapts the global cycle length $t_c(k_n)$ based on the maximum observed degree of saturation within the network:

\begin{equation}
d_{\max}(k_n) = \max_{n \in \mathcal{N}_g} \max_{z \in I_n} d_z(k_n).
\end{equation}

If the network is highly saturated,
\begin{equation}
d_{\max}(k_n) \geq d_{\text{upper}},
\end{equation}
the cycle length is increased proportionally:

\begin{equation}
t_c(k_{n+1}) = \min \left( t_c(k_n) + \alpha_c (d_{\max} - d_{\text{upper}}),\; t_{c,\max} \right).
\end{equation}

If the network is underutilised,
\begin{equation}
0 < d_{\max}(k_n) < d_{\text{lower}},
\end{equation}
the cycle length is decreased:

\begin{equation}
t_c(k_{n+1}) = \max \left( t_c(k_n) - \alpha_c (d_{\text{lower}} - d_{\max}),\; t_{c,\min} \right).
\end{equation}

Otherwise, the current cycle length is retained.
This gradual adaptation corresponds to the SCOOT principle of maintaining operation near a target degree of saturation.

% --------------------------------------------------
\textit{2. Green Split Optimisation:}

Given the updated cycle length, the effective green time at intersection $n$ is:

\begin{equation}
t_{\text{eff},n} = t_c - |F_n| \cdot t_L.
\end{equation}

Green splits are adapted based on the most critical (highest saturated) lane at each intersection.

If the corresponding queue length exceeds a threshold, additional green time is allocated to the phase serving $z^{\star} = \arg\max_{z \in I_n} d_z(k_n)$. 
The increment is proportional to the difference in degree of saturation between the most critical and competing phases:

\begin{equation}
g_{n,j^{\star}}(k_{n+1}) =
g_{n,j^{\star}}(k_n)
+ \alpha_g \cdot \Delta d,
\end{equation}

subject to upper bounds (e.g., limiting dominance to three quarters of the effective cycle).

The remaining green time is redistributed among the other phases proportionally or through normalisation to ensure:

\begin{equation}
\sum_{j \in F_n} g_{n,j}(k_{n+1}) = t_{\text{eff},n}.
\end{equation}

If no lane exceeds the activation threshold, green splits are scaled proportionally to the updated cycle length.
Thus, green split optimisation is intersection-local but dependent on the globally coordinated cycle length.

% --------------------------------------------------
\textit{3. Offset Optimisation:}
The final step coordinates intersections along corridors by adjusting offsets $\theta_n$ to promote progression (``green waves'').

First, district-level congestion is estimated by normalising cumulative queue lengths by total lane length within each district, as exemplified in Figure~\ref{fig:scosca}. 
The district with the highest normalised congestion is identified as critical.

Travel times between consecutive intersections are estimated as:

\begin{equation}
\tau_{n \rightarrow m} = \frac{L_{n \rightarrow m}}{v_{\text{limit}}},
\end{equation}

where $L_{n \rightarrow m}$ is the total connection length and $v_{\text{limit}}$ the assumed progression speed.

If the congestion gap between districts exceeds a predefined threshold, offsets are adjusted sequentially along the critical corridor:

\begin{equation}
\theta_{m}(k_{n+1}) =
\min \left(
\theta_{n}(k_n) + \alpha_o \tau_{n \rightarrow m},
\; t_c
\right),
\end{equation}

thereby aligning downstream green onsets with upstream vehicle arrivals.

A hysteresis band is introduced to avoid oscillatory behaviour when congestion differences are small.

\subsubsection{Exclusion of Other Controllers}

% Numerous signal control approaches have been proposed in the literature, ranging from fixed-time optimisation methods (e.g.\ Webster-based design), actuated control, dynamic programming approaches such as OPAC and RHODES, model predictive control (MPC), and more recently reinforcement learning-based strategies.

% The objective of this chapter, however, is not to provide an exhaustive benchmarking of all available signal control algorithms, but to compare representative control paradigms within a unified and structurally transparent modelling framework.

The two selected controllers — Max-Pressure and SCOOT/SCATS — span the principal dimensions of signalised intersection management, and suffice for this work's goal of computational benchmarks:

\begin{itemize}
    \item Decentralised, state-feedback, queue-based optimisation (Max-Pressure),
    \item Hierarchical, cycle-based, coordinated adaptive control (SCOOT/SCATS).
\end{itemize}

Together, these controllers represent the two dominant philosophies in urban signal control:
reactive pressure-based allocation of right-of-way and coordinated cycle-based progression control.

Other approaches fall largely into refinements or hybridisations of these paradigms. 
Fixed-time and Webster-type designs~\cite{Webster1958} do not incorporate real-time feedback and therefore do not provide a meaningful benchmark within a dynamic simulation environment. 
Fully actuated controllers~\cite{newell1960queues} represent threshold-based variants of local feedback control and are structurally subsumed by pressure-based formulations. 
Dynamic programming and model predictive control (MPC-based) approaches~\cite{de1998optimal} require explicit traffic state prediction models and substantial calibration effort, which would introduce additional modelling layers beyond the scope of this study. 
Reinforcement learning methods~\cite{bakker2010traffic}, while promising, rely on extensive training procedures and hyper-parameter tuning, making performance comparisons highly dependent on scenario-specific configuration rather than control structure.

By restricting the analysis to Max-Pressure and Scoot/Scats, the study captures both decentralised optimal control and coordinated adaptive network control while maintaining interpretability, reproducibility, and methodological consistency. 
Additional controllers would primarily increase algorithmic diversity without expanding the underlying control-theoretic spectrum represented in the comparison.

% #######################################################################################
% #######################################################################################
%\newpage
\section{Guidelines For Simulation-Design and Sensor Placement} \label{guideline}

In this section, we present essential guidelines for simulation design and sensor placement to enhance the realism of SUMO simulations and ensure that vehicle behaviour aligns with the intended operational objectives.
While SUMO's default parameters work well with urban networks, several specific adjustments are recommended for freeway networks, as discussed below.

\subsection{Case Study for Ramp Metering}

\begin{figure}[!h]
    \centering
    \includegraphics[width=\linewidth]{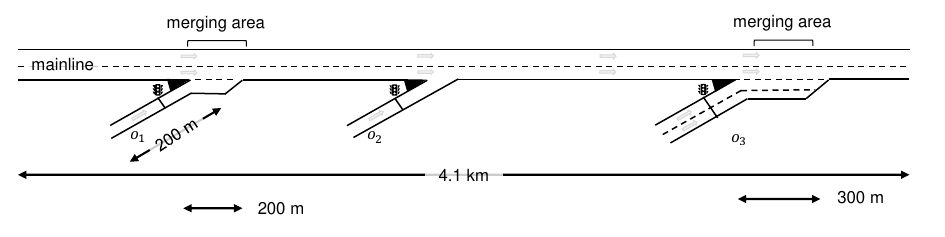}
    \caption{Ramp Metering Case Study.}
    \label{fig:case_study_ramp_metering}
\end{figure}

The ramp metering case study considers a two-lane motorway section of length 4.1\,km, equipped with three metered on-ramps, each with a ramp length of approximately 200\,m, as shown in Figure~\ref{fig:case_study_ramp_metering}. 
In order to analyse the influence of geometric design on merging behaviour and controller performance, three distinct on-ramp configurations are investigated. 
In Version~1, the on-ramp consists of a single lane and is complemented by an auxiliary merge lane on the mainline of approximately 200\,m, allowing vehicles to accelerate and merge more gradually. 
In Version~2, the on-ramp is also single-lane but does not provide an additional merge lane on the motorway, requiring vehicles to merge directly into the rightmost mainline lane. 
Version~3 represents a higher-capacity configuration with multiple ramp lanes and an extended  auxiliary merge lane of approximately 300\,m on the mainline. 
These geometric variations enable a systematic evaluation of how merging conditions and available weaving space affect congestion formation, queue dynamics, and the performance of the applied ramp metering strategies.
    
\FloatBarrier

\subsubsection{Simulation Model Design}

A dedicated warm-up period should precede the actual analysis interval. 
By initializing the simulation earlier than the measurement start time, the network can reach representative traffic conditions, preventing distortions in performance indicators caused by unrealistically low initial traffic densities.
Furthermore, the simulation time step should be chosen sufficiently small to ensure an accurate and numerically stable representation of vehicle dynamics and interactions. 
A finer temporal resolution improves the modelling of car-following, lane-changing, and merging behaviour, particularly in congested or highly dynamic traffic situations.

In this case study, a total simulation duration of 4,200 s was selected to represent one hour of analysis time (08:00-09:00 a.m.), preceded by a 10-minute warm-up period. The simulation time step was set to 0.5,s per step. 
Based on our experience, this value represents the coarsest time resolution that still provides stable and sufficiently accurate traffic dynamics while maintaining computational efficiency.

\subsubsection{Network Model Design}

The design of the network model is a critical component of highway simulations in SUMO. 
\textbf{Particular care is required when importing motorway geometries from OpenStreetMap}, as automatically generated networks frequently contain geometric inconsistencies, unrealistic junction configurations, or improper lane connections.

In many cases, \textbf{it can be more efficient and reliable to manually construct the motorway infrastructure} using the Netedit tool (with satellite imagery as background reference) rather than correcting numerous artefacts from imported map data. \textbf{A carefully designed network ensures realistic merging behaviour, correct lane usage, and stable traffic dynamics.} Also speed limits should be double checked on highway edges to ensure realistic velocity conditions.

\begin{figure}[!h]
    \centering
    \includegraphics[width=\linewidth]{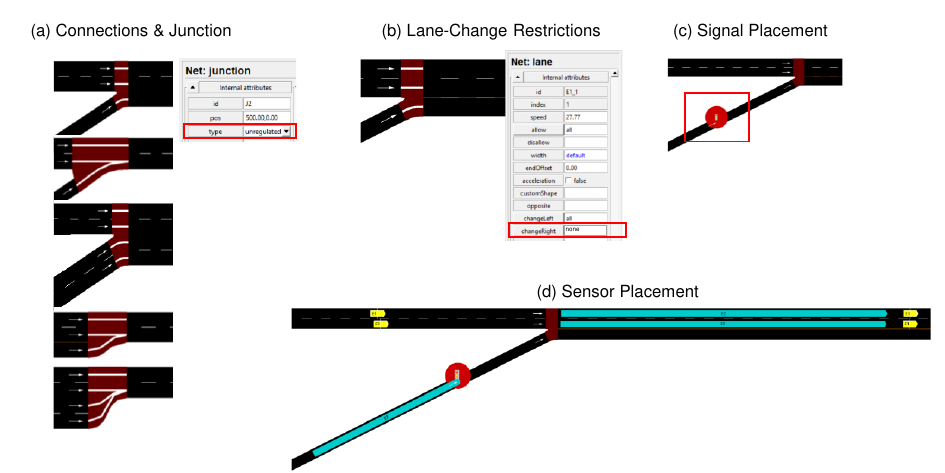}
    \caption{\textbf{Freeway Network Model Design Considerations.}}
    \label{fig:case_study_ramp_metering_net}
\end{figure}
\FloatBarrier

Merging operations at on-ramps are substantially affected~\footnote{More details can be found here: \url{https://sumo.dlr.de/docs/Simulation/Motorways.html\#motorway\_ramps}.} by (i) junction type, (ii) connection configuration at the end of the auxiliary lane and (iii) signal and (iv) sensor placement, which should be considered during network construction and are discussed in the following:

\paragraph{Connections and Merging Geometry}

All lane-to-lane connections must be explicitly verified to ensure realistic merging behaviour. 
Incorrect or overly permissive connections (such as U-turns) can lead to unintended lane-changing patterns or unrealistic vehicle trajectories. 
Special attention should be paid to merge areas, ensuring that ramp vehicles are properly connected to the intended mainline lanes.
See Figure~\ref{fig:case_study_ramp_metering_net}(a) as a reference.

\paragraph{Junction Type Configuration}

The junction type at merge points strongly influences driver behaviour. 
For on-ramp merging areas, unregulated junctions (i.e., priority-based merges rather than signal-controlled or all-way regulated intersections) are typically more appropriate. 
If the junction is incorrectly configured as regulated, ramp vehicles may behave overly cautiously and wait excessively for large gaps, thereby underestimating merging pressure and its impact on mainline congestion. 
Proper priority settings ensure that ramp vehicles merge assertively and realistically influence motorway traffic.
See Figure~\ref{fig:case_study_ramp_metering_net}(a) as a reference.

\paragraph{Mainline Lane-Change Restrictions}

In scenarios with auxiliary merge lanes, it is important to restrict unrealistic lane-changing behaviour. 
Specifically, mainline vehicles should be prevented from changing into the additional merge lane that is intended exclusively for ramp traffic. 
Without such restrictions (e.g., forbidding \texttt{changeRight} where appropriate), mainline vehicles may occupy the auxiliary lane, which rarely occurs in real-world motorway operations and can significantly distort congestion patterns.
See Figure~\ref{fig:case_study_ramp_metering_net}(b) as a reference.

\paragraph{Ramp Metering Signal Infrastructure}

When modelling ramp metering, the on-ramp edge should be split to create a dedicated intersection for the traffic signal.
Sufficient downstream distance must be provided between the traffic light and the physical merge point to allow vehicles to accelerate before entering the mainline. 
If this acceleration section is too short, vehicles will merge at unrealistically low speeds, artificially reducing mainline capacity and exaggerating congestion effects.
See Figure~\ref{fig:case_study_ramp_metering_net}(c) as a reference.

\paragraph{Sensor Placement}

Accurate performance assessment and controller implementation require carefully positioned detectors.
See Figure~\ref{fig:case_study_ramp_metering_net}(d) as a reference.

\begin{itemize}
    \item \textit{Ramp detectors:}
Lane area detectors (E2) are recommended upstream of the ramp signal to measure queue length and occupancy. A detector length of approximately 50,m has proven suitable for capturing ramp queues while maintaining sufficient resolution for control purposes.
    \item \textit{Mainline detectors:}
Detectors should be placed on each mainline lane but never on auxiliary merging lanes, as this would bias flow and density measurements. Depending on the evaluation objective, detectors may be located:
\begin{itemize}
    \item Upstream of the merge area (e.g., 10,m before the merge),
    \item Within the merging section (e.g., centrally located),
    \item Downstream of the merge (e.g., 100,m after the merge).
\end{itemize}

Both induction loop detectors (E1) and lane area detectors (E2) can be used, depending on whether point-based flow measurements or spatial occupancy and density estimates are required.
\end{itemize}

% There are many things to consider when creating network, especially when highway network model is copied from Open Street Map.
% Sometimes it can be faster to draw the highway yourself with satellite image in background of Netedit software rather than to clean all the flaws from the open street map model.

% Certain things to consider:
% 	- Connection Mode: check conncetions make sense for merging
%     - Intersection / Junction type: unregulated (visible through white crossing connections rather than some gray)
%     otherwise, ramp users not aggressively enough enter highway and always wait, instead of causing congestion on mainline by entering in
%     - Mainline: forbid ChangeRight for the mainline lanes intersecting with extra merge lane
% 					(visible through orange line), otherwise some cars from mainline even make lane changes to the additional merging lane that is actually intended for the ramp users (this would most certainly never happen in reality)
%     - Signals: on-ramp split edge for extra intersection with traffic light (have enough space after ramp meter for vehicles to accelerate again otherwise they are too slow when entering the highway at the on-ramp intersection
%     - Sensors: 
%         Ramp Sensors: laneAreaDetector (E2) for queue length (e.g. 50m long) in front of traffic light, 
%         Mainline Sensors (never on merging lane(s)): 
%             on each lane, can be upstream (e.g. 10 meters), donwstream (e.g. 100m), or in merging area (e.g. in middle)
%             can be inductionLoop (E1) or laneAreaDetector (E2)

\subsubsection{Demand Model Design}

A carefully designed demand model is essential for obtaining realistic traffic dynamics in highway simulations using SUMO. 
In particular, both fleet composition and vehicle insertion mechanisms strongly influence congestion formation, merging behaviour, and overall network performance.

\paragraph{Fleet Composition}

For motorway scenarios, realistic congestion patterns require heterogeneous driving behaviour within the vehicle population. Homogeneous driver settings often lead to overly stable traffic flow and underestimate the formation of disturbances and capacity drops.

Based on empirical experience, multiple (at least five), distinct driver behaviour groups should be defined for each vehicle category (e.g., passenger cars and motorcycles) and for each origin (mainline and on-ramp). 
Mainline vehicles should be parametrised to behave more cooperatively, whereas ramp vehicles should be configured to exhibit slightly more assertive behaviour in order to realistically represent merging dynamics.

The demand generation process therefore includes:
\begin{itemize}
    \item Definition of routes for each origin (mainline and ramps),
    \item Specification of multiple driving behaviour parameter sets per vehicle type and origin,
    \item Equal splitting of each vehicle flow among the aggressiveness levels,
    \item Use of \texttt{vTypeDistribution} to assign behavioural probabilities.
\end{itemize}

This structured heterogeneity enables more realistic lane-changing manoeuvrers, merging conflicts, and shock-wave formation, particularly under high demand conditions.

In this case study, we define five aggressivity levels for passenger cars and motorcycles (vehicle types), for each origin, as summarised in Table~\ref{tab:aggressiveness_levels}.

\begin{table}[!h]
    \centering
    \caption{\textbf{Heterogeneous Driving Behaviour Parameters.} Definition of aggressiveness levels for mainline and ramp vehicles (passenger cars and motorcycles). Values shown represent systematic trends from Level~1 (least aggressive) to Level~5 (most aggressive).}
    \label{tab:aggressiveness_levels}
    \small
    \begin{tabular}{llccccc}
        \toprule
        \textbf{Origin} & \textbf{Parameter} & \textbf{Level 1} & \textbf{Level 2} & \textbf{Level 3} & \textbf{Level 4} & \textbf{Level 5} \\
        \midrule
        \multirow{6}{*}{Mainline}
        & speedFactor      & 0.80 & 0.85 & 0.90 & 0.95 & 1.00 \\
        & $\tau$ [s]       & 10.0 & 9.8  & 9.3  & 8.8  & 8.5  \\
        & lcCooperative    & 1.0  & 1.0  & 1.0  & 1.0  & 1.0  \\
        & lcAssertive      & 0.01 & 0.05 & 0.08 & 0.10 & 0.20 \\
        & accel [m/s$^2$]  & 0.5  & 0.7  & 0.8  & 1.0  & 1.1  \\
        & decel [m/s$^2$]  & 3.5  & 4.0  & 4.5  & 5.0  & 5.5  \\
        \midrule
        \multirow{6}{*}{Ramp}
        & speedFactor      & 0.80 & 0.85 & 0.90 & 0.95 & 1.00 \\
        & $\tau$ [s]       & 1.5  & 0.9  & 0.8  & 0.6  & 0.4  \\
        & lcCooperative    & 0.03 & 0.01 & 0.00 & 0.00 & 0.00 \\
        & lcAssertive      & 9.5  & 10.3 & 10.8 & 11.3 & 12.0 \\
        & accel [m/s$^2$]  & 4.0  & 4.3  & 4.6  & 4.8  & 5.0  \\
        & decel [m/s$^2$]  & 3.0  & 4.0  & 5.0  & 6.0  & 8.0  \\
        \bottomrule
    \end{tabular}
\end{table}

\paragraph{Flow Definition and Vehicle Spawning}

The vehicle insertion mechanism has a substantial impact on achieved traffic flow. 
In highway simulations, we strongly recommend to generate vehicles using a probabilistic (geometrically distributed) departure process rather than fixed, periodic departure (e.g., via \texttt{vehsPerHour}). 
A probabilistic spawning process better reflects real-world traffic arrivals and avoids artificial platooning effects caused by equidistant vehicle insertions.

It is important to note that the configured demand in the route file does not necessarily correspond to the actually realised flow in SUMO. 
Due to default safety checks and insertion constraints, the number of vehicles successfully inserted into the network can be significantly lower than specified. 
Therefore, \textbf{flows should always be verified using detectors or via TraCI-based measurements} instead of assuming that the nominal demand equals the effective traffic flow.

To achieve realistic and sufficiently high motorway flows, the following insertion parameters are recommended:
\begin{itemize}
    \item \texttt{probability} instead of \texttt{vehsPerHour}
(The probability value is defined per second; \texttt{probability = 1.0} corresponds to 3600 veh/h.)
    \item \texttt{departSpeed="max"} to avoid unrealistic standstill departures on highways,
    \item \texttt{departLane="free"} to allow flexible lane assignment at insertion,
    \item \texttt{departPos="last"} to distribute vehicles along the entry edge and prevent artificial congestion at the network boundary,
    \item \texttt{insertionChecks="none"} to reduce overly restrictive insertion behaviour that can suppress achievable flow.
\end{itemize}

Even when applying these configurations, the effective flow is still limited by network conditions and SUMO’s internal constraints. 
Consequently, it is strongly recommended to validate the realised traffic volumes through detector-based measurements or TraCI queries to ensure consistency between intended and simulated demand levels.

\subsection{Case Study for Intersection Management}

\begin{figure}[!h]
    \centering
    \includegraphics[width=\linewidth]{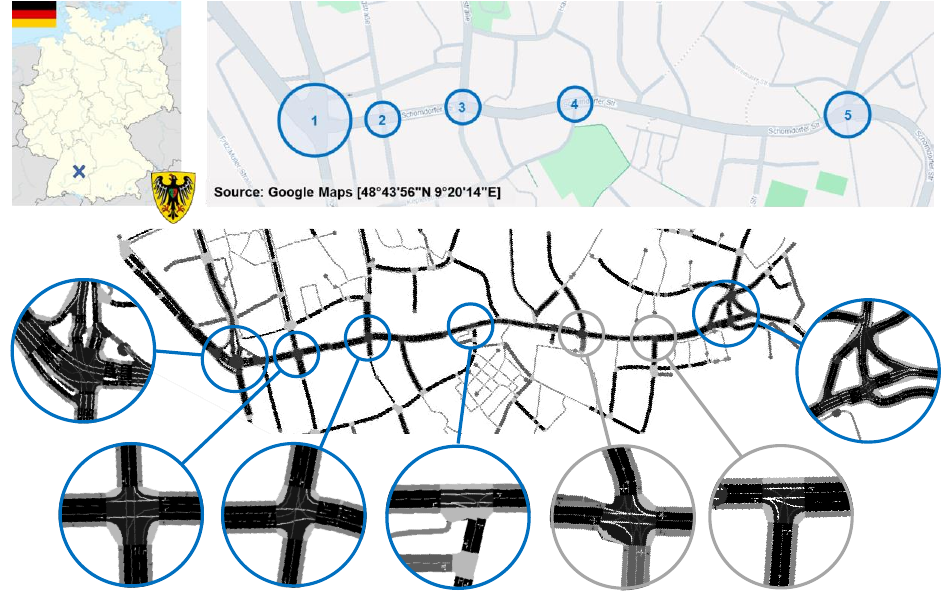}
    \caption{Signalised Intersection Management Case Study.}
    \label{fig:case_study_signal_man}
\end{figure}

The signalised intersection management case study is taken from~\cite{riehl2025green}, which is a demand-calibrated microsimulation\footnote{The demand model that was calibrated using loop-detector data, and originates from a cooperation with the municipal administration Tiefbauamt Esslingen am Neckar (Germany).} of an arterial network (see Figure~\ref{fig:case_study_signal_man}), and comprises seven signalised intersections with more than 96 considered traffic light signals\footnote{For legal reasons, our model only covers the control of five out of the seven intersections.}.

\FloatBarrier

\subsubsection{Simulation Model Design}

A dedicated warm-up period should precede the actual analysis interval also in the urban context. 
Furthermore, the simulation time step should be chosen sufficiently small to ensure an accurate and numerically stable representation of vehicle dynamics and interactions. 
Especially in urban context, choosing a small simulation time step of 0.25 s per step turned out to be most suitable to capture the microscopic vehicle dynamics.

The simulation runs on a usual work day from 09:00 am to 24:00 pm (on March 4th, 2024), and was executed at time steps of 0.25 s.

\subsubsection{Network Model Design}

In addition to the aspects mentioned in context of freeway network design, a crucial aspect in urban network design is the definition of movement phases at intersections and sensor placement for signalised intersection management.

\begin{figure}
    \centering
    \includegraphics[width=\linewidth]{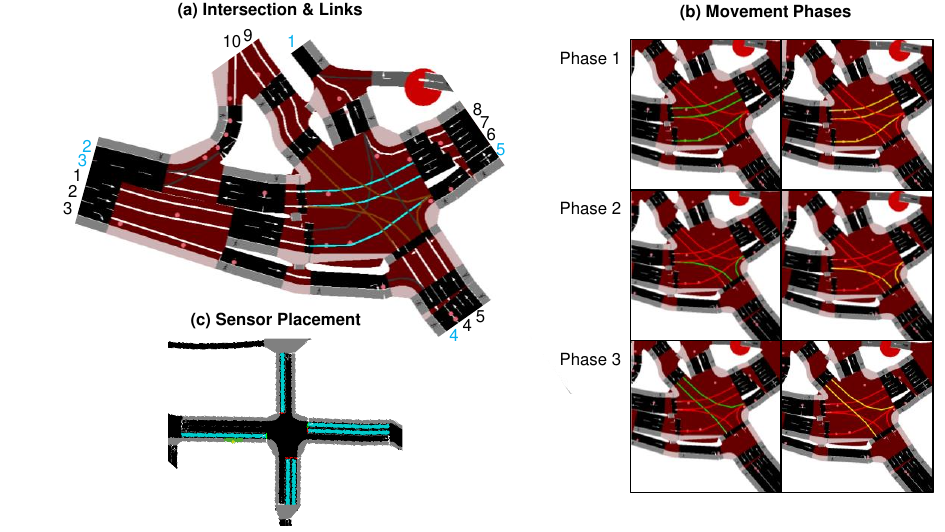}
    \caption{\textbf{Urban Network Model Design Considerations.}}
    \label{fig:netedit_inter}
\end{figure}

\paragraph{Intersection Design}

At each signalised intersection, one needs to define all links (approaches) and combine those to movement phases (lane-to-lane connections), as shown in Figure~\ref{fig:netedit_inter}. 
Subfigure (a) shows one exemplary intersection, and all ingoing links (black numbers) and outgoing links (blue numbers). 
Subfigure (b) shows their combination to movement phases (links that simultaneously achieve way of right.

\paragraph{Sensor Placement}

In terms of sensor placement, each of the links require separate sensors for queue length and density measurements.
This can be achieve through placing lane area sensors (E2) at each link, as shown in Subfigure (c).
It is important to leave enough space for the sensors to enable a fair queue measurement due to equal sensor lengths.
Furthermore, in the implementation of \texttt{sumoITScontrol}, we enable not only to use sensors but also direct access to information from lanes themselves (list of lanes or sensors related to a specific link).
Access to the lanes themselves is simpler implementation-wise, and could be assumed to be realised with camera sensors in practice.
The queue lengths of a movement phase corresponds to the aggregated queue length of the subset of all links that are entering at the specific phase.

\subsubsection{Demand Model Design}

Contrary to the freeway case study, no differentiation in different aggressivity classes was necessary to reproduce urban congestion dynamics.
The simulation includes around 26,714 single vehicle trips through the network, from 24 origins to 18 destinations.
In addition to the vehicle trips, 11 bus lines that pass 26 bus stops ($20$ s stop time each) in the arterial network and surrounding are modelled according to the time schedule, resulting in a total of 48 additional public transport trips.
Following the vehicle fleet composition statistics of Germany~\cite{fahrzeugbestand} provided by the Kraftfahrt-Bundesamt (Federal Office of Transport of Germany), 81\% of all single vehicle trips are conducted by passenger-cars, 8.2\% by motorcycles, 4.6\% by transporters, and 6.2\% by trucks. 
Moreover, 2\% of the passenger-cars are considered to be electric vehicles.

% #######################################################################################
% #######################################################################################
%\newpage
\section{Best-Practices for Stochastic Calibration \& Evaluation} \label{bestpract}

\subsection{Considerations For Stochastic Optimisation}

\textit{Stochasticity of Microsimulations:} In contrast to analytical, macroscopic traffic flow models, microscopic traffic simulation models such as \texttt{SUMO} exhibit inherent stochasticity arising from driver behaviour models, lane-changing decisions, departure processes, and interaction effects. 
As a result, simulation outcomes are subject to non-negligible variability even when model parameters remain unchanged.

\textit{Multiple Runs Required:} Consequently, single-run experiments are insufficient to derive statistically meaningful conclusions regarding controller performance. 
Instead, multiple independent simulation runs with different random seeds must be conducted. 
This procedure enables the estimation of central tendency (e.g., mean performance) and dispersion (e.g., standard deviation or confidence intervals), thereby allowing robust comparison between control strategies and against an uncontrolled baseline scenario.
Neglecting replication may lead to overfitting controllers to specific random realizations, misinterpretation of improvements, and limited reproducibility of results. 
A rigorous stochastic evaluation framework is thus essential for scientifically sound controller assessment.
At the same time, given the computational cost of microscopic simulations, an appropriate trade-off must be found between statistical reliability and computational effort, as too few runs undermine validity while excessive replications may become prohibitively expensive. \textbf{We recommend at least 10-20 repetitions.}

\textit{Reporting Performance \textbf{AND} Variance:} We therefore strongly recommend performing a sufficiently large number of replications for each experimental configuration. 
Performance indicators should be reported together with measures of variability, and, where appropriate, formal statistical hypothesis testing should be applied to assess whether observed differences are statistically significant rather than artefacts of stochastic fluctuations.
Furthermore, beyond variance measures such as standard deviation, it is highly recommended to report 95\% confidence intervals, box-plots or distribution visualisations, and effect sizes.

\textit{Paired Experimental Design:} When comparing two controllers, it is advisable to evaluate them under identical stochastic realizations (i.e., identical random seeds). 
This paired experimental design reduces variance in the comparison and increases statistical power.

\textit{Documentation of Computational Experiments:} To ensure reproducibility and transparency, computational experiments must be documented in a comprehensive and structured manner~\cite{riehl2025revisiting}. 
In stochastic microscopic simulations, reproducibility depends not only on model parameters, but also on random seeds, simulation duration, step size, demand specifications, detector configurations, and software versions. 
Even minor changes in these settings may alter the results.
We therefore strongly recommend reporting all relevant simulation parameters, including the list of random seeds used for evaluation. 
Given the structural complexity of microscopic traffic simulations, it is best practice to share the complete experimental setup, including network files, demand definitions, configuration files, and controller implementations. 
Providing open-source access (e.g., via a public GitHub repository) substantially improves reproducibility, facilitates peer verification, and strengthens scientific credibility. 
Without transparent documentation, independent replication of results is often impractically difficult.

\textit{Statistical Significance Evaluation with Tests:} Beyond reporting mean performance and variance, formal statistical hypothesis testing should be conducted to assess whether observed differences between controllers (ideally improvements) are statistically significant. 
When performance metrics are approximately normally distributed and paired experiments are used (i.e., identical random seeds across controllers), a paired t-test~\cite{student1908ttest,ross2014introprobstat} provides a powerful method for detecting significant differences in mean performance. 
If normality cannot be assumed, non-parametric alternatives such as the Wilcoxon signed-rank test~\cite{wilcoxon1945signedrank} (for paired samples) or the Mann-Whitney U test~\cite{mann1947mannwhitney} (for independent samples) may be applied. 
Prior to selecting a test, it is advisable to examine the empirical distribution of the performance metric using normality diagnostics or visual inspection. 
In addition, when comparing multiple controllers simultaneously, appropriate corrections for multiple hypothesis testing should be considered to control the overall error rate.
The use of formal statistical tests strengthens the validity of conclusions and prevents over-interpretation of stochastic fluctuations as genuine performance improvements.

% Contrary to analytic, macroscopic simulation models, microscopic simulation models such as SUMO are more turbulent and noisy, therefore not only one but multiple experiments need to be conducted to make statistically meaningful statements on the efficiency and improvement of controllers relative to each other and to the uncontrolled scenario.

% We highly recommend to multiple SUMO simulations with different random seeds, and to regard mean and standard deviation to make informed decisions and enable a statistically valid analysis of the performance of controllers.

\subsection{Control Parameter Calibration with Stochastic Optimisation}

Calibrating control parameters in a stochastic microscopic simulation environment requires a structured optimisation framework. 
In general, three key components must be defined: (i) the parameter space, (ii) the objective (goal) function, and (iii) the stochastic sampling strategy.
By explicitly defining these three components, stochastic controller calibration becomes a well-posed optimisation problem rather than an ad-hoc parameter search. 
This structured approach improves robustness, transparency, and reproducibility of the calibration process.

\textit{Definition of the Parameter Space:} 
First, a set of controller parameters to be calibrated must be specified together with their admissible ranges. 
These ranges may be derived from theoretical considerations, engineering experience, or preliminary sensitivity analyses. 
The search space might be discrete or continuos.
Clearly defining the search space prevents unrealistic parameter configurations and ensures comparability across experiments. 
Depending on the controller complexity, the parameter space may be low-dimensional (e.g., a single gain parameter) or multi-dimensional (e.g., gains, thresholds, cycle times).
The more control parameters can be (assumed to) fixed, the lower the computational efforts for calibrating the variable, to-be-identified parameters.
Depending on the size and continuity of the parameter space, it can make sense to either conduct a grid-search of optimal parameters, or to apply any other optimisation algorithms.

\textit{Definition of the Objective Function:} 
Second, a goal function must be formulated to quantify controller performance. 
This function translates system-level performance indicators (e.g., total time spent, delay, queue length, throughput, occupancy deviations) into a scalar evaluation metric. 
In many practical applications, the objective function consists of a weighted combination of multiple performance measures to balance competing objectives such as efficiency, stability, and fairness. 
The choice of weights should be justified, as it directly influences the optimisation outcome.

\textit{Definition of the Stochastic Sampling Strategy:} 
Third, the stochastic nature of microscopic simulation must be explicitly integrated into the optimisation procedure. 
For each candidate parameter configuration, performance should be evaluated over a predefined set of random seeds. 
The optimisation target is therefore not a single simulation outcome, but the expected performance (e.g., mean objective value) across replications. 
Using a fixed and documented set of seeds across parameter configurations ensures comparability and reduces evaluation noise. 
In this context, the optimisation problem becomes the minimisation (or maximisation) of the expected objective value under stochastic variability.

\subsection{Exemplary Fine Tuning of ALINEA}

In this section, we demonstrate the importance of stochastic evaluation by calibrating the parameters of the ALINEA ramp metering controller within the considered case study using stochastic, computational optimisation. 

\textit{Parameter Space:} To reduce complexity and focus on methodological aspects, we assume a simplified ALINEA formulation using a proportional (P) control law only. 
The proportional gain parameter $K_P$ is optimized within a predefined search space ranging from 5 to 50. 
All remaining controller parameters are kept constant throughout the experiment, including the target occupancy, control cycle duration, and minimum and maximum metering rates. 
This allows us to isolate the effect of the gain parameter on overall system performance.

\textit{Objective Function:}
The objective function we want ALINEA to be optimised for is the weighted sum of the average queue length over the whole simulation run (weight=1) and the average occupancy violation (weight=5), where the violation refers to the clipped difference between observed occupancy and target occupancy.

\textit{Stochastic Sampling Strategy:}
For each value of $K_P$ in the search space, 20 independent simulation runs with different random seeds are conducted. 
For each configuration (parameter set), the mean performance and standard deviation are computed. 
The aggregated results are summarized in Table~\ref{tab:alinea_tuning}.

\begin{table}[h]
    \centering
    \caption{\textbf{Stochastic Optimisation for ALINEA Calibration.}}
    \label{tab:alinea_tuning}
    \begin{tabular}{c c c}
    \hline
    Configuration ($K_P$) & Mean Performance & Standard Deviation \\
    \hline
    5  & 13.041 & 2.438 \\
    10 & 12.481 & 2.278 \\
    15 & 12.894 & 2.368 \\
    20 & 12.637 & 2.343 \\
    25 & 13.355 & 2.580 \\
    30 & 13.320 & 2.649 \\
    35 & 13.758 & 2.694 \\
    40 & 13.839 & 2.699 \\
    45 & 14.166 & 2.697 \\
    50 & 13.982 & 2.698 \\
    \hline
    \end{tabular}
\end{table}

The results reveal a non-monotonic relationship between the proportional gain and system performance. 
The lowest mean performance value is obtained for $K_P = 10$, with a mean of 12.48 and a standard deviation of 2.28. 
However, neighbouring parameter values (e.g., $K_P = 15$ or $K_P = 20$) exhibit similar mean performance within the range of stochastic variability. 
For larger gain values ($K_P \geq 35$), both the mean performance and the variability increase, indicating reduced robustness and potential overreaction of the controller.

Importantly, the observed differences between parameter settings are small relative to their standard deviations. 
Without multiple replications, a single simulation run could easily lead to a different ranking of parameter configurations. 
This example therefore illustrates that reliable controller calibration in microscopic traffic simulation requires replicated experiments and variance-aware evaluation, rather than relying on single-run outcomes.

\paragraph{Statistical Significance Analysis:} 

Is the controller with configuration $K_P=10$ statistically significantly better than the configuration with $K_P=5$? 

In the following analysis we are trying to answer this question by conducting a two-sample t-test.

First, we have two averages $\bar{x}_5=13.041$, and $\bar{x}_{10}=12.481$, two standard deviations $s_5=2.438$ and $s_{10}=2.278$, and two sample sizes $n_5=20$ and $n_{10}=20$ (as both configurations were evaluated for 20 times each).

Second, we try to test whether the null hypothesis can be falsified or not: $H_0: \mu_{10} = \mu_5$ meaning there is no significant improvement of configuration 10 over 5.

Third, the t-statistic is calculated as follows:
\begin{equation}
    t = \frac{\Delta}{SE} = \frac{\bar{x}_5-\bar{x}_{10}}{s_p\sqrt{\frac{1}{n_5}+\frac{1}{n_{10}}}} \approx 0.75
\end{equation} 
with $s_p$ being the pooled standard deviation:
\begin{equation}
    s_p = \sqrt{\frac{(n_5-1)s_5^2 + (n_{10}-1)s_{10}^2}{n_5+n_{10} - 2}}.
\end{equation}

Fourth, looking up the two-sided p-value for the given test statistic $t$ results in $p=0.46$ (one-sided $p=0.23$), meaning, that the \textbf{improvement of configuration 10 over 5 is} \textbf{not statistically significant} at a common significance level $\alpha=0.05$.
The stochastic variability is of similar magnitude as the performance differences between candidate parameter settings, and therefore statistically grounded evaluation is indispensable.

\subsection{Further Readings}
There is a multitude of further guidelines that can be used as a reference, and are mentioned here for completeness, such as \textit{Hinweise zur mikroskopischen Verkehrsflusssimulation}~\cite{fgsv} or \textit{Traffic Analysis Toolbox Volume VI}~\cite{dowling2007traffic}.

% In this section we calibrate (fine tune) the parameters of the ALINEA controller at the given case study for ramp metering.
% To simplify, we assume an ALINEA controller with just a proportional P-contorl law, and optimize its parameter in a search space from 5-50.
% All other parameters are fixed (e.g. target occupancy, cycle duration, min and maximum metering rate).
% For each parameter in the parameter space we conduct 20 experiments.
% The results are shown in table below (mean, std),
% 5 [13.041207561889655, 2.4384136099917487]
% 10 [12.481044009619161, 2.2775428870985557]
% 15 [12.894482957963698, 2.3681340957184524]
% 20 [12.636514045816126, 2.342537274130512]
% 25 [13.35457413165786, 2.5804242493244146]
% 30 [13.319907248459165, 2.6485499285459422]
% 35 [13.757885740292059, 2.6943198053512983]
% 40 [13.838714296130345, 2.6991941556358597]
% 45 [14.165949450774553, 2.69682343548254]
% 50 [13.982141129842612, 2.6979434403798134]

% #######################################################################################
% #######################################################################################
%\newpage
\section{Conclusion} \label{conclusion}

This paper introduced \texttt{sumoITScontrol}, an open-source and extensible Python framework providing a curated collection of widely used traffic controllers implemented for the microscopic simulation environment SUMO via the \texttt{TraCI} interface. In contrast to proposing yet another novel control strategy, the primary objective of this work was to address a methodological gap in the literature: the lack of transparent, standardised, and reproducible baseline implementations for benchmarking traffic control algorithms under consistent microscopic simulation settings.

The framework integrates established controllers for both freeway and urban traffic management, including ramp metering approaches such as \texttt{ALINEA}, \texttt{METALINE}, and \texttt{HERO}, as well as signal control strategies such as \texttt{Max-Pressure} and \texttt{SCOOT/SCATS}. 
By translating these classical, theory-driven methods into modular and well-documented implementations, \texttt{sumoITScontrol} lowers the barrier to rigorous experimental comparison and reduces the need for repeated, project-specific re-implementation of baseline algorithms.

Beyond software provision, this paper emphasised methodological best-practices for simulation design, sensor placement, stochastic calibration, and statistical evaluation. 
In microscopic traffic simulation, stochastic variability arising from driver behaviour models, lane-changing decisions, and departure processes can substantially influence performance metrics. 
Through the exemplary calibration of the \texttt{ALINEA} controller, we demonstrated that performance differences between parameter configurations may be small relative to their standard deviations, and that single-run evaluations can easily lead to misleading conclusions. 
The results underline the necessity of replicated experiments, variance-aware reporting, and formal statistical hypothesis testing to ensure scientifically sound controller assessment.

The presented case studies for freeway ramp metering and urban intersection control illustrate how the framework can be employed to design reproducible experiments, conduct structured parameter optimisation, and compare controllers under controlled stochastic conditions. 
In doing so, this work contributes not only a software package, but also a reproducibility-oriented experimental paradigm for ITS research in microscopic environments.

Nevertheless, several limitations remain. First, the provided controller set, while extensive, is not exhaustive. 
Emerging learning-based and connected-vehicle-oriented strategies are not yet systematically integrated. Second, the case studies focus on representative but specific network configurations; broader validation across diverse traffic scenarios, demand patterns, and network topologies would further strengthen generalisability. 
Third, although stochastic calibration and evaluation guidelines are provided, automated large-scale benchmarking pipelines could further enhance usability and comparability.

Future work will therefore focus on extending the controller library to variable speed limit and perimeter control, and integrating additional adaptive and learning-based approaches, and developing automated benchmarking workflows with standardised evaluation protocols. 
Furthermore, incorporating realistic sensor noise models, communication delays, and mixed traffic compositions (including connected and electric vehicles) would enable more comprehensive assessments of controller robustness under near-realistic operating conditions.

In summary, \texttt{sumoITScontrol} establishes a transparent and reproducible foundation for benchmarking traffic control strategies in SUMO. 
By combining open-source implementations with methodological guidance on stochastic optimisation and statistical evaluation, the framework aims to strengthen experimental standards, improve comparability across studies, and foster more robust scientific progress within the ITS and SUMO research communities.

% #######################################################################################
% #######################################################################################
\clearpage
\section*{Data availability statement}
The source code, examples, and documentation can be found on the project's GitHub page \url{https://github.com/DerKevinRiehl/sumoITScontrol/} and the pip/PyPi archive page \url{https://pypi.org/project/sumoITScontrol/}.

% \section*{Underlying and related material}
% If you have other material which supports your findings (e.g. model code) or is closely related to your article/contribution (e.g. supplementary material as videos, samples, etc.) deposited on a repository, please include a brief description and the respective DOI(s) here.

\section*{Author contributions}
\begin{itemize}
    \item \textbf{Kevin Riehl:} Conceptualisation, Methodology, Software, Validation, Investigation, Resources, Writing - Original Draft, Visualisation, Project administration, Funding acquisition.
    \item \textbf{Anastasios Kouvelas:} Writing - Review \& Editing, Supervision.
    \item \textbf{Michail A. Makridis:} Writing - Review \& Editing, Supervision, Funding acquisition.
\end{itemize}

\section*{Competing interests}
The authors declare that they have no competing interests.

\section*{Funding}
This work has received funding from the European Union’s Horizon Europe research and innovation programme under grant agreement No. 101203465 (project "FEDORA"), and from the Swiss State Secretariat for Education, Research and Innovation (SERI) under contract No. REF-1131-52301.

\section*{Acknowledgements}
{\emergencystretch=3em
This work contributes to REproducible Research In Transportation Engineering (\href{https://www.rerite.org/}{RERITE}), advancing open science and transparency in transportation research, and the vision of a transportation community where open science, reproducible research and replicable studies are the norms, advancing scientific rigour, accelerating innovation and fostering translation into practice.
}

\section*{Legal Disclosure on Commercial Traffic Controllers}
The authors of this work herewith confirm and emphasize, that the implementations of the here-mentioned controllers correspond to publicly available information as references in this work.
The authors do not claim that their implementation equals or discloses copyrighted implementation details of their commercial counterparts, this we explicitly state for Max-Pressure, SCOOTs, SCATS, and HERO.

\clearpage
\printbibliography[heading=references]

% #######################################################################################
% #######################################################################################
\newpage
\appendix
\renewcommand{\thesubsection}{A.\arabic{subsection}}
\section{Appendix: Demonstration \& Results}
%\section*{Appendix: Demonstration \& Results}
%\section{Demonstrations \& Results} \label{res}

\subsection{Ramp Metering}

\subsubsection{ALINEA}

\begin{figure} [!h]
    \centering
    \includegraphics[width=1.0\linewidth]{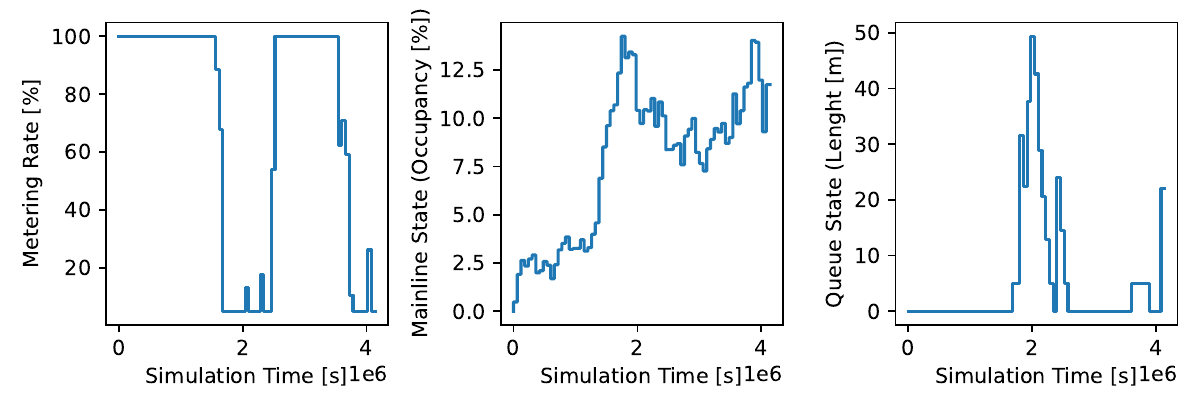}
    \includegraphics[width=1.0\linewidth]{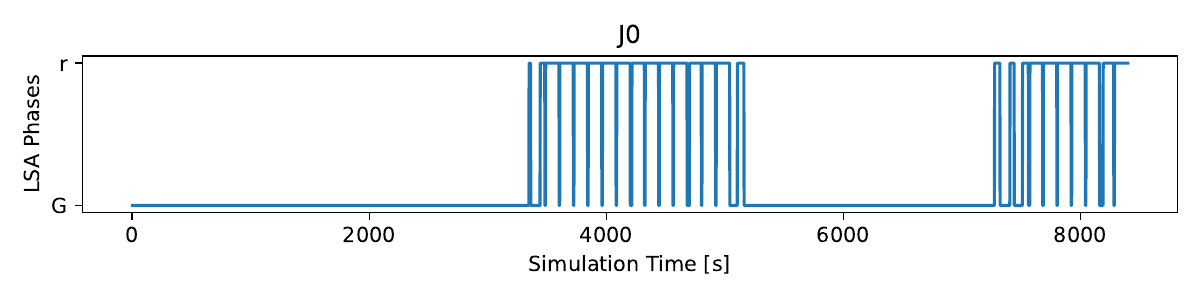}
    \caption{\textbf{Demonstration ALINEA.}}
    \label{fig:demo_alinea}
\end{figure}

Figure~\ref{fig:demo_alinea} shows the control actions (metering rate), measured system state (mainline occupancy, queue length) and traffic signals.
From the following parametrization, one can see that it effectively achieves to stabilize the occupancy around 10\% (its target occupancy) once demand rises, while this causes a queue to grow on the ramp.

{\small
\begin{verbatim}
    ramp_meter = RampMeter(
        tl_id="J0",
        mainline_sensors=["e2_5", "e2_4"],
        queue_sensors=["e2_0"],
    )
    controller = ALINEA(
        params={
            "target_occupancy": 10,
            "K_P": 30,
            "K_I": 0,
            "cycle_duration": 60,
            "measurement_period": int(
                60 / 0.5
            ),  # int(cycle_duration / simulation.time_step)
            "min_rate": 5,
            "max_rate": 100,
        },
        ramp_meter=ramp_meter,
    )
\end{verbatim}
}

\FloatBarrier

\FloatBarrier

\newpage
\subsubsection{METALINE}

\begin{figure} [!h]
    \centering
    \includegraphics[width=1.0\linewidth]{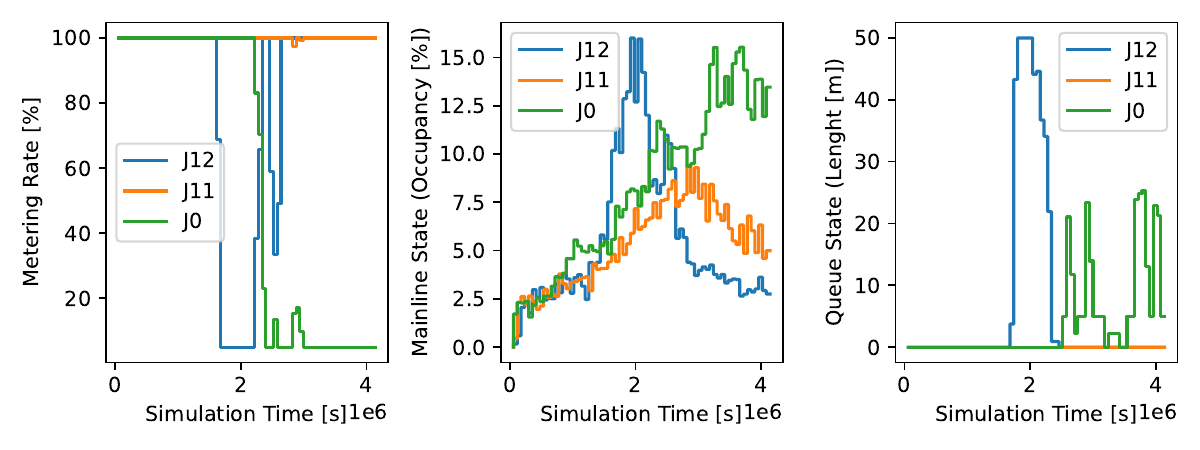}
    \includegraphics[width=1.0\linewidth]{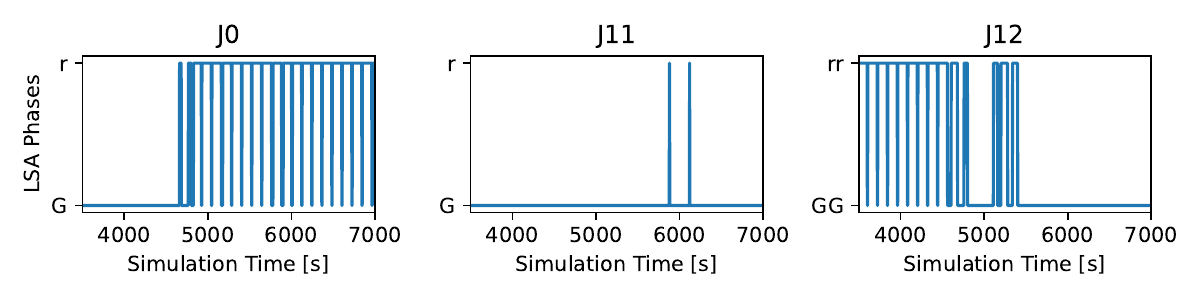}
    \caption{\textbf{Demonstration METALINE.}}
    \label{fig:demo_metaline}
\end{figure}

Figure~\ref{fig:demo_metaline} shows the control actions (metering rates), measured system state (mainline occupancy, queue length) and traffic signals, for three coordinated ramp meters.
From the following parametrization, one can see that keeping occupancy stable around target was not working too well for all ramps.
Stronger stabilisation on some ramps (e.g. J11) came at the cost of more instability of other ramps (J0, J12) or excessive queue forming.
\texttt{METALINE} was not optimised further, which explains its suboptimal behaviour.
This demonstrates the complexity of finding optimal parameters for \texttt{METALINE}.

{\small
\begin{verbatim}
    ramp_meter_group = RampMeterCoordinationGroup(
        ramp_meters_ordered=[
            RampMeter(
                tl_id="J12",
                mainline_sensors=["e1_13", "e1_14"],
                queue_sensors=["e2_1", "e2_2"],
                smoothening_factor=0.1,
                saturation_flow_veh_per_sec=0.5,
            ),
            RampMeter(
                tl_id="J11",
                mainline_sensors=["e1_2", "e1_3"],
                queue_sensors=["e2_3"],
                smoothening_factor=0.1,
                saturation_flow_veh_per_sec=0.5,
            ),
            RampMeter(
                tl_id="J0",
                mainline_sensors=["e2_5", "e2_4"],
                queue_sensors=["e2_0"],
                smoothening_factor=0.1,
                saturation_flow_veh_per_sec=0.5,
            ),
        ],
        ramp_meter_ids=["J12", "J11", "J0"],
    )
    controller = METALINE(
        params={
            "cycle_duration": 60,  # control cycle
            "measurement_period": int(
                60 / 0.5
            ),  # int(cycle_duration / simulation.time_step)
            "min_rate": 5,
            "max_rate": 100,
        },
        coordination_group=ramp_meter_group,
        target_occupancies=[10, 10, 10],
        # Interaction gain matrix (3x3)
        K_P=np.array(
            [
                [30, -5, 0],   # ramp 1 influenced negatively by ramp 2
                [-3, 25, -2],  # ramp 2 influenced by neighbors
                [0, -4, 20],
            ]
        ),
        # Optional integral gain matrix
        K_I=np.zeros(shape=(3, 3)),
    )
\end{verbatim}
}

\FloatBarrier

\FloatBarrier

\newpage
\subsubsection{HERO}

\begin{figure} [!h]
    \centering
    \includegraphics[width=1.0\linewidth]{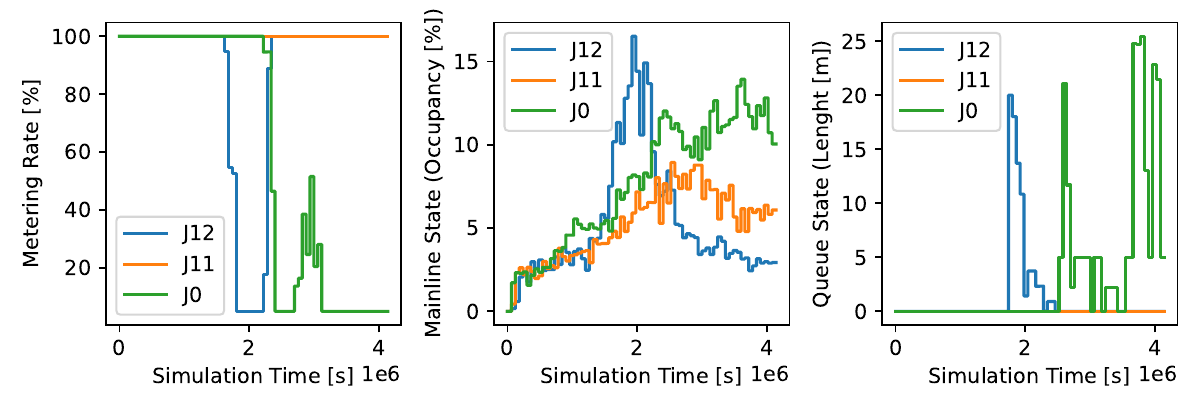}
    \includegraphics[width=1.0\linewidth]{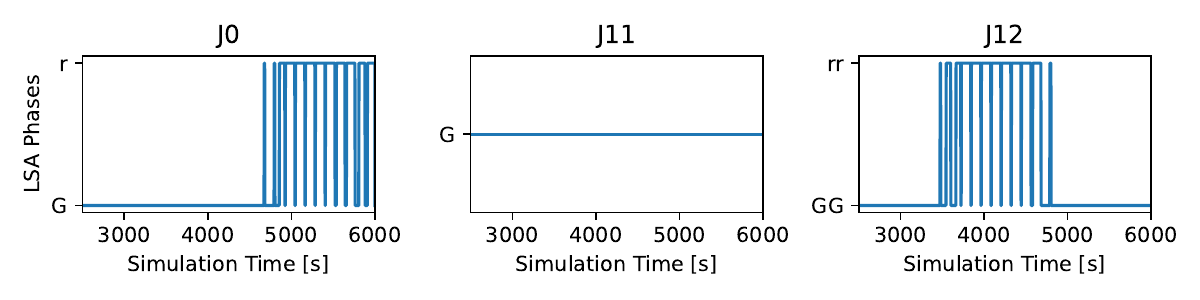}
    \includegraphics[width=1.0\linewidth]{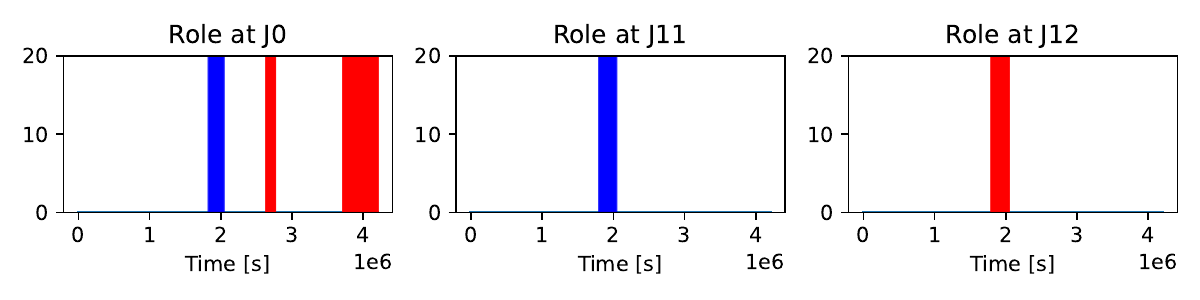}
    \caption{\textbf{Demonstration HERO.}}
    \label{fig:demo_hero}
\end{figure}

Figure~\ref{fig:demo_hero} shows the control actions (metering rates), measured system state (mainline occupancy, queue length) and traffic signals, for three coordinated ramp meters.
In addition to that (in the bottom row of plots) it demonstrates which ramp becomes a master (red) and slave (blue) over the time of the simulation.
Compared to \texttt{METALINE}, \texttt{HERO} is able to improve the situation for ramp J0.

{\small
\begin{verbatim}
    ramp_meter_group = RampMeterCoordinationGroup(
        ramp_meters_ordered=[
            RampMeter(
                tl_id="J12",
                mainline_sensors=["e1_13", "e1_14"],
                queue_sensors=["e2_1", "e2_2"],
                smoothening_factor=0.1,
                saturation_flow_veh_per_sec=0.5,
            ),
            RampMeter(
                tl_id="J11",
                mainline_sensors=["e1_2", "e1_3"],
                queue_sensors=["e2_3"],
                smoothening_factor=0.1,
                saturation_flow_veh_per_sec=0.5,
            ),
            RampMeter(
                tl_id="J0",
                mainline_sensors=["e2_5", "e2_4"],
                queue_sensors=["e2_0"],
                smoothening_factor=0.1,
                saturation_flow_veh_per_sec=0.5,
            ),
        ],
        ramp_meter_ids=["J12", "J11", "J0"],
    )
    controller = HERO(
        params={
            "hero_period": 60,  # similar to ALINEA cycle duration
            "queue_activation_threshold_m": 15.0,  # master queue trigger
            "queue_release_threshold_m": 2.5,  # dissolve cluster
            "min_queue_setpoint_m": 5.0,  # for slaves
            "anticipation_factor": 1.0,  # factor to obtain nonconservative ...
            # ... prediction of demand to come in next control period
            "avg_vehicle_spacing": 7.5,  # average vehicle spacing to convert ...
            # ... meters to vehicles and vice versa, from queue length measurements
        },
        coordination_group=ramp_meter_group,
        alinea_controllers={
            "J12": ALINEA(
                params={
                    "target_occupancy": 10,
                    "K_P": 30,
                    "K_I": 0,
                    "cycle_duration": 60,
                    "measurement_period": int(
                        60 / 0.5
                    ),  # int(cycle_duration / simulation.time_step)
                    "min_rate": 5,
                    "max_rate": 100,
                },
                ramp_meter=ramp_meter_group.ramp_meters[0],
            ),
            "J11": ALINEA(
                params={
                    "target_occupancy": 10,
                    "K_P": 30,
                    "K_I": 0,
                    "cycle_duration": 60,
                    "measurement_period": int(
                        60 / 0.5
                    ),  # int(cycle_duration / simulation.time_step)
                    "min_rate": 5,
                    "max_rate": 100,
                },
                ramp_meter=ramp_meter_group.ramp_meters[1],
            ),
            "J0": ALINEA(
                params={
                    "target_occupancy": 10,
                    "K_P": 30,
                    "K_I": 0,
                    "cycle_duration": 60,
                    "measurement_period": int(
                        60 / 0.5
                    ),  # int(cycle_duration / simulation.time_step)
                    "min_rate": 5,
                    "max_rate": 100,
                },
                ramp_meter=ramp_meter_group.ramp_meters[2],
            ),
        },
    )
\end{verbatim}
}

\FloatBarrier

\FloatBarrier

\newpage 

\subsection{Signalised Intersection Management}

\subsubsection{Max-Pressure (Fixed)}

\begin{figure} [!h]
    \centering
    \includegraphics[width=0.7\linewidth]{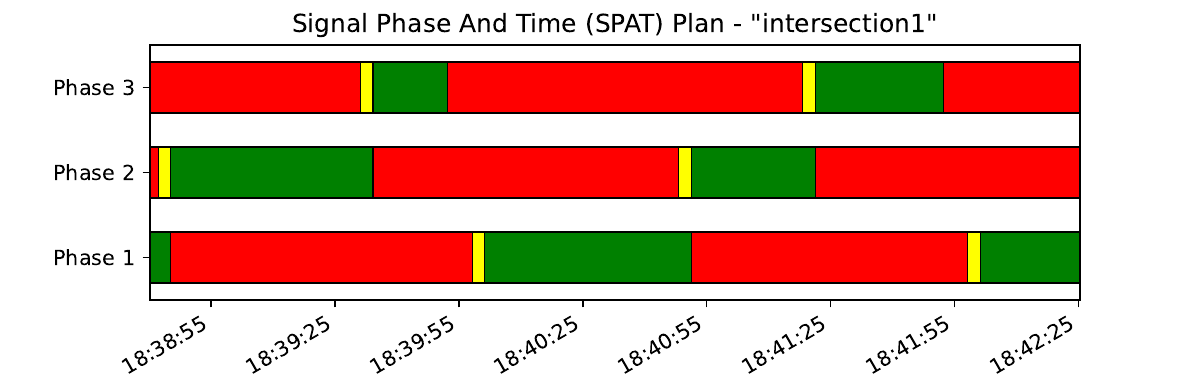}
    \includegraphics[width=0.7\linewidth]{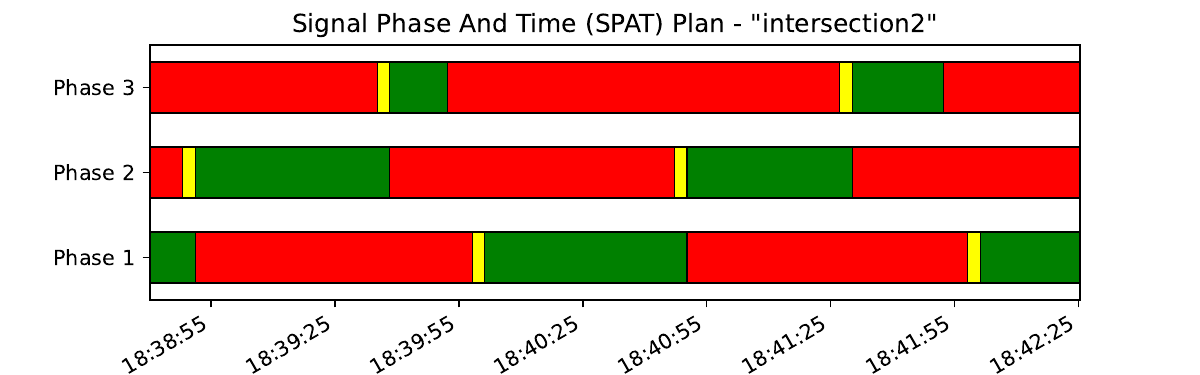}
    \includegraphics[width=0.7\linewidth]{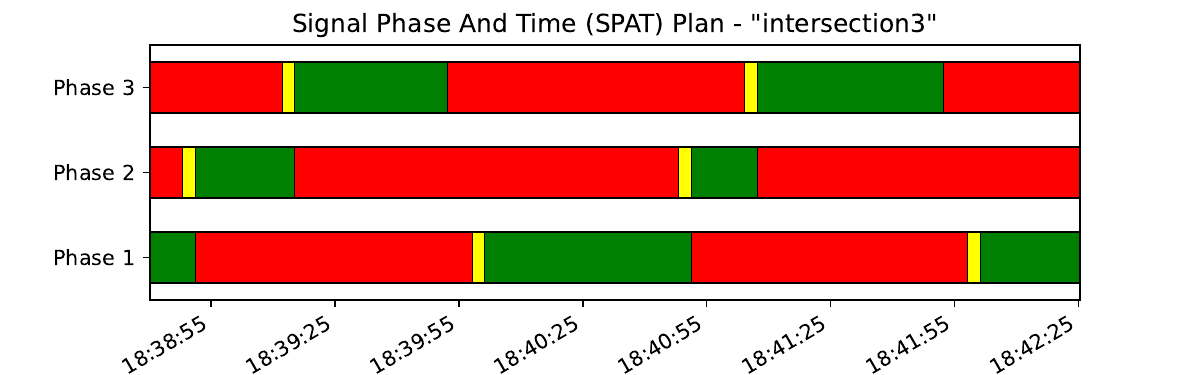}
    \includegraphics[width=0.7\linewidth]{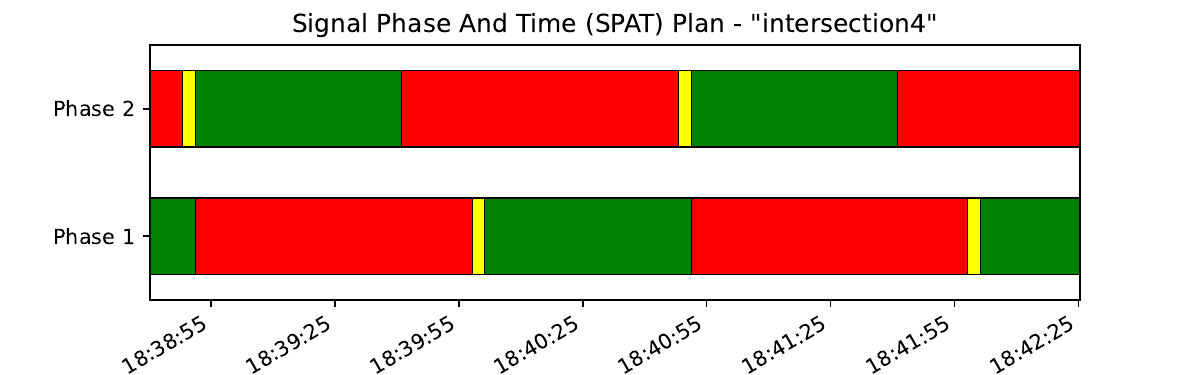}
    \includegraphics[width=0.7\linewidth]{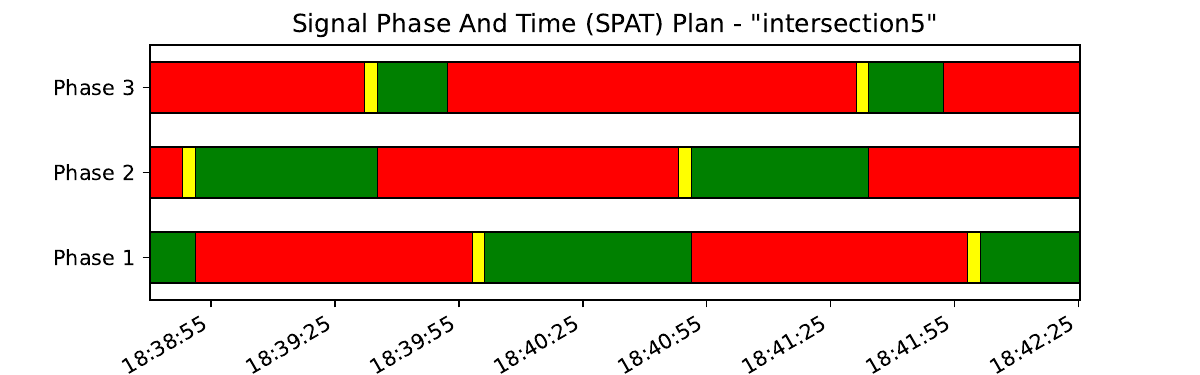}
    \caption{\textbf{Demonstration Max-Pressure (Fixed) (1/2).}}
    \label{fig:demo_maxfix}
\end{figure}

\begin{figure} [!h]
    \centering
    \includegraphics[width=1.0\linewidth]{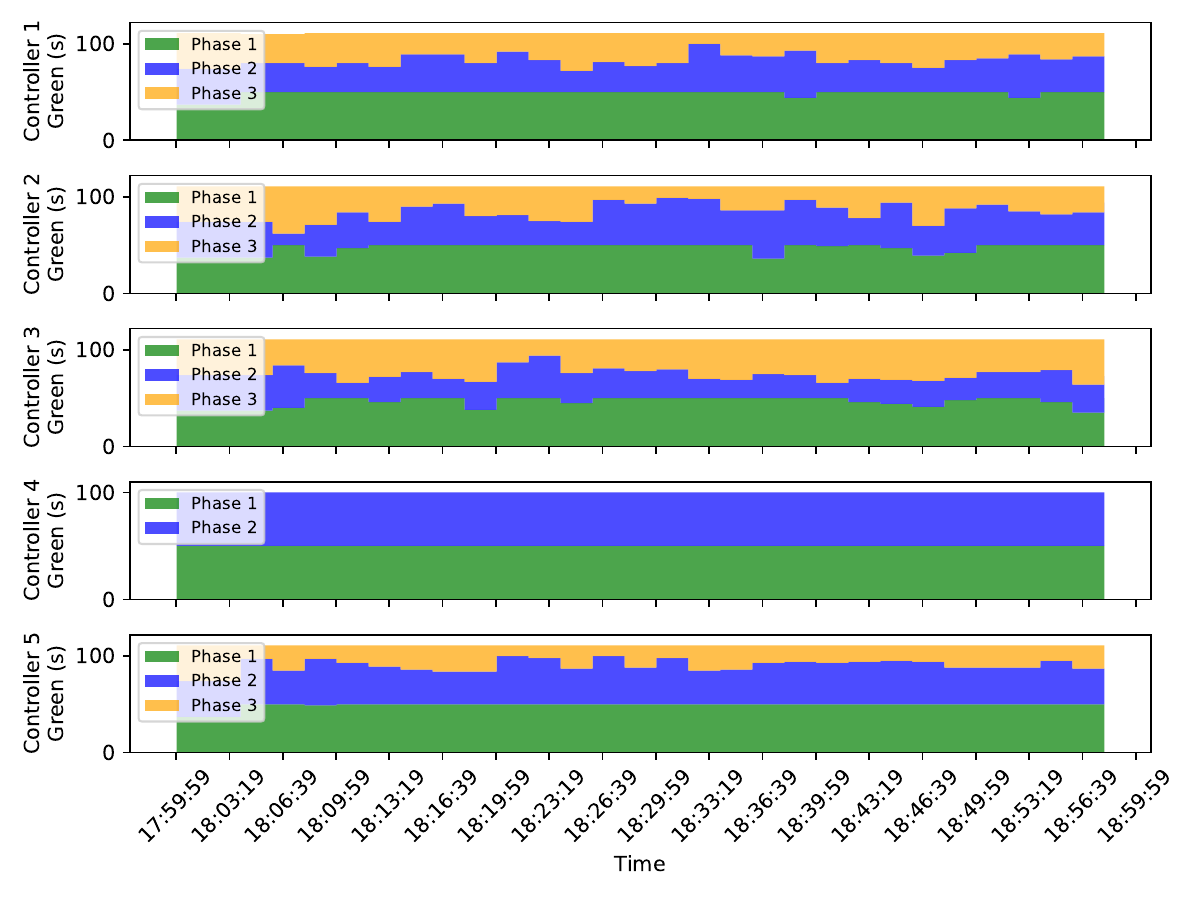}
    \caption{\textbf{Demonstration Max-Pressure (Fixed) (2/2).}}
    \label{fig:demo_maxfix2}
\end{figure}

Figure~\ref{fig:demo_maxfix} shows the signal phase and time (SPAT) plans for all four intersections (zoomed in for 3.5 minutes).
The order of phases is fix (always starting from Phase 1, then 2, and then 3).
Figure~\ref{fig:demo_maxfix2} shows the signal schedules over time.
The cycle duration (sum of all phases' green times) is constant, but the split across the phases changes.

{\small
\begin{verbatim}
    intersection1 = Intersection(
        tl_id="intersection1",
        phases=[0, 2, 4],
        links={
            0: ["921020465#1_3","921020465#1_2","921020464#0_1","921020464#1_1",
                "38361907_3","38361907_2","-1164287131#1_3","-1164287131#1_2", ],
            2: ["-1169441386_2","-1169441386_1","-331752492#1_2","-331752492#1_1",
                "-331752492#0_1","-331752492#0_2",],
            4: ["-183419042#1_1","26249185#30_1","26249185#30_2","26249185#1_1",
                "26249185#1_2",],
        },
        green_states=["GGrrGrr", "rrGGrrr", "rrrrrGG"],
        yellow_states=["yyrryrr", "rryyrrr", "rrrrryy"],
    )
    intersection2 = Intersection(
        tl_id="intersection2",
        phases=[0, 2, 4],
        # links = {0:["183049933#0_1", "-38361908#1_1"],
        #           2:["-38361908#1_1", "-38361908#1_2"],
        #           4:["-25973410#1_1", "758088375#0_1", "758088375#0_2"]},
        sensors={
            0: ["e2_183049933#0_1", "e2_-38361908#1_1"],
            2: ["e2_-38361908#1_1", "e2_-38361908#1_2"],
            4: ["e2_-25973410#1", "e2_758088375#0_1", "e2_758088375#0_2"],
        },
        green_states=["GGrrrGrr", "GGGrrrrr", "rrrGGrGG"],
        yellow_states=["yyrrryrr", "yyyrrrrr", "rrryyryy"],
    )
    intersection3 = Intersection(
        tl_id="intersection3",
        phases=[0, 2, 4],
        links={
            0: ["E3_1", "-758088377#1_1", "-758088377#1_2", "-E1_1", "-E1_2"],
            2: ["E3_1", "E3_2"],
            4: ["-758088377#1_1", "-E1_1", "-E4_1", "-E4_2"],
        },
        green_states=["GGGrrrr", "rrGGrrr", "GrrrGGG"],
        yellow_states=["yyyrrrr", "rryyrrr", "yrrryyy"],
    )
    intersection4 = Intersection(
        tl_id="intersection4",
        phases=[0, 2],
        links={
            0: ["22889927#0_1", "758088377#2_1", "-22889927#2_1"], 
            2: ["-25576697#0_0"]
        },
        green_states=["GgrrGG", "rrGGGr"],
        yellow_states=["yyrryy", "rryyyr"],
    )
    intersection5 = Intersection(
        tl_id="intersection5",
        phases=[0, 2, 4],
        links={
            0: ["E6_1", "E6_2", "E5_1", "130569446_1", "E15_1", "E15_2"],
            2: ["E15_2", "E6_3", "E5_2", "130569446_2"],
            4: ["E10_1", "E9_1", "1162834479#1_1", "-208691154#0_1", "-208691154#1_1"],
        },
        green_states=["GGrrrrGGrrrr", "rrGrrrrrGrrr", "rrrGGgrrrGGg"],
        yellow_states=["yyrrrryyrrrr", "rryrrrrryrrr", "rrryyyrrryyy"],
    )
    
    max_pressure_params = {
        "T_L": 3,  # Yellow Time
        "G_T_MIN": 5,  # Min Greentime (used for Max. Pressure)
        "G_T_MAX": 50,  # Max Greentime (used for Max. Pressure)
        "measurement_period": int(1 / 0.25),  # int(1 / simulation.time_step)
        "cycle_duration": 120,
    }
    controller1 = MaxPressure_Fix(params=max_pressure_params, intersection=intersection1)
    controller2 = MaxPressure_Fix(params=max_pressure_params, intersection=intersection2)
    controller3 = MaxPressure_Fix(params=max_pressure_params, intersection=intersection3)
    controller4 = MaxPressure_Fix(params=max_pressure_params, intersection=intersection4)
    controller5 = MaxPressure_Fix(params=max_pressure_params, intersection=intersection5)
\end{verbatim}
}

\FloatBarrier

\FloatBarrier

\newpage

\subsubsection{Max-Pressure (Flexible)}

\begin{figure} [!h]
    \centering
    \includegraphics[width=0.7\linewidth]{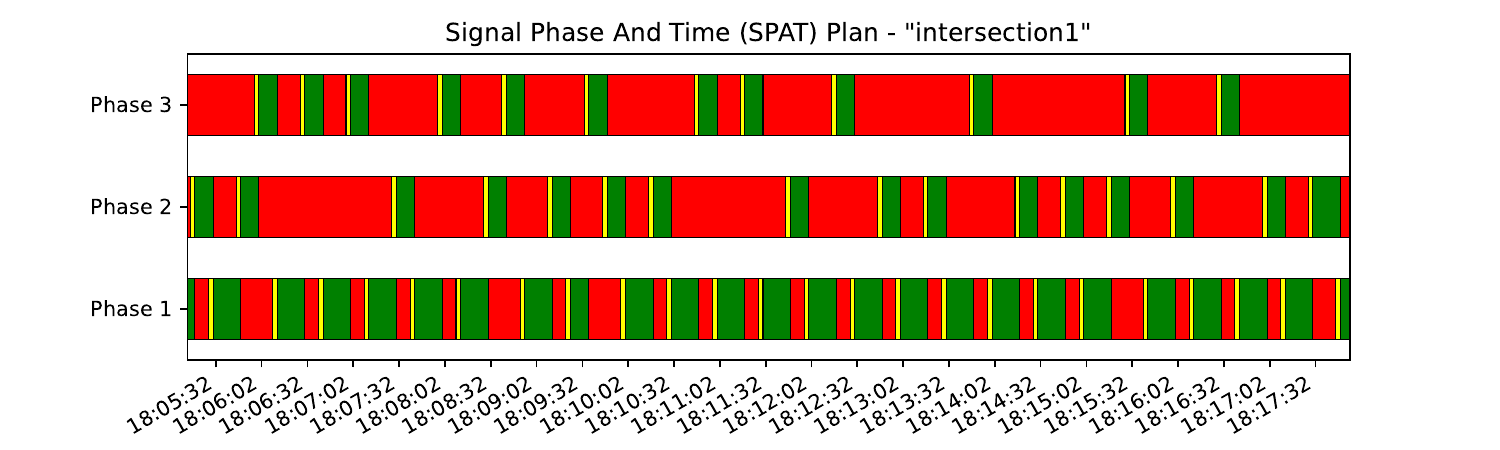}
    \includegraphics[width=0.7\linewidth]{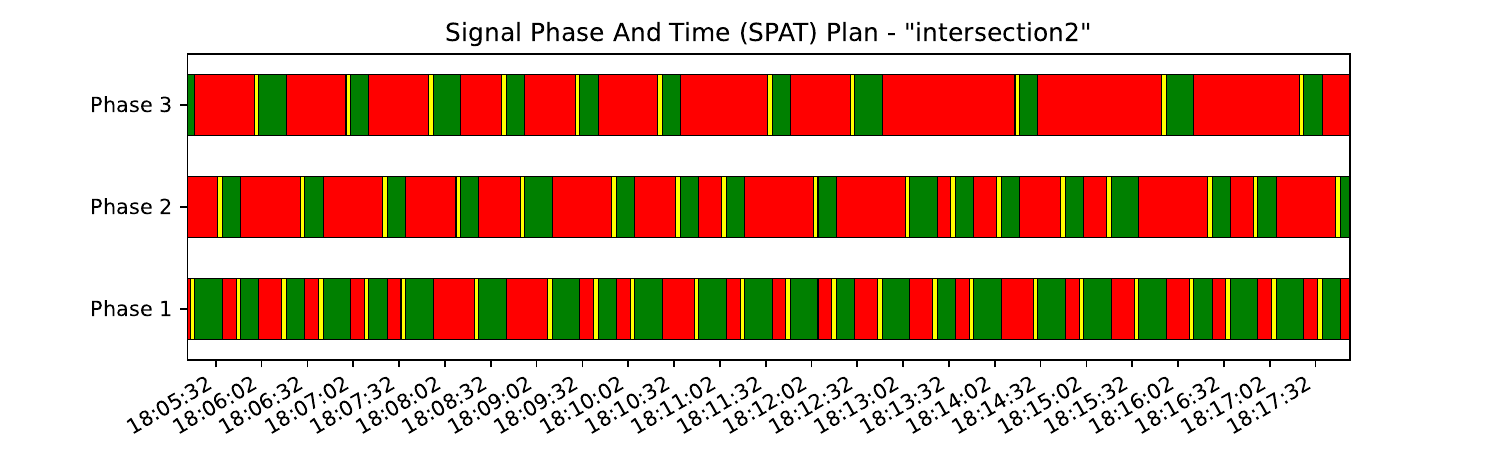}
    \includegraphics[width=0.7\linewidth]{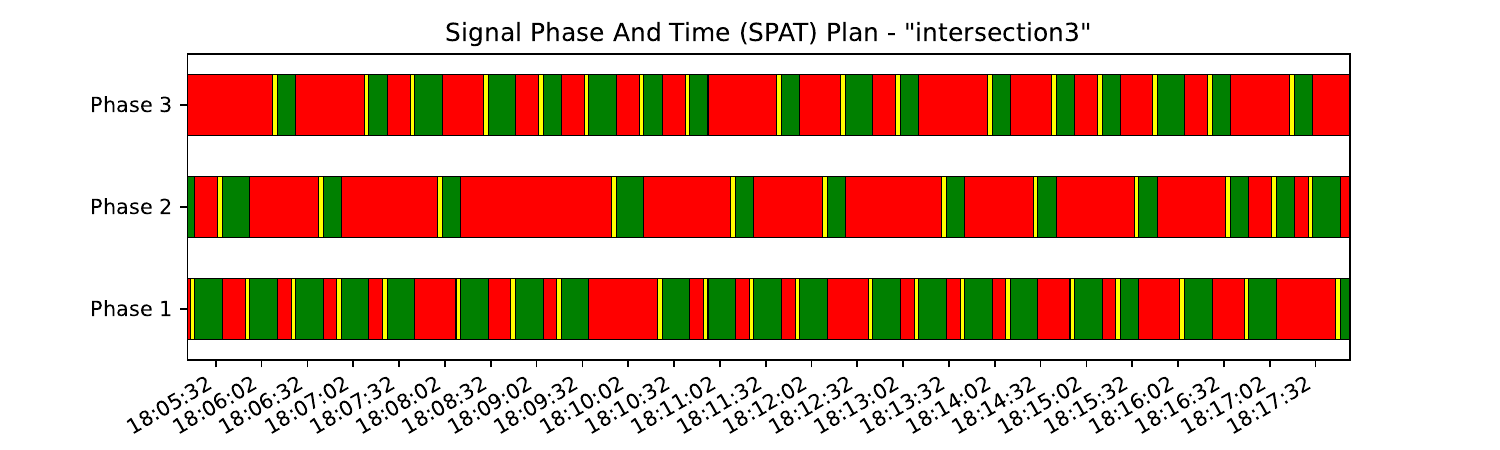}
    \includegraphics[width=0.7\linewidth]{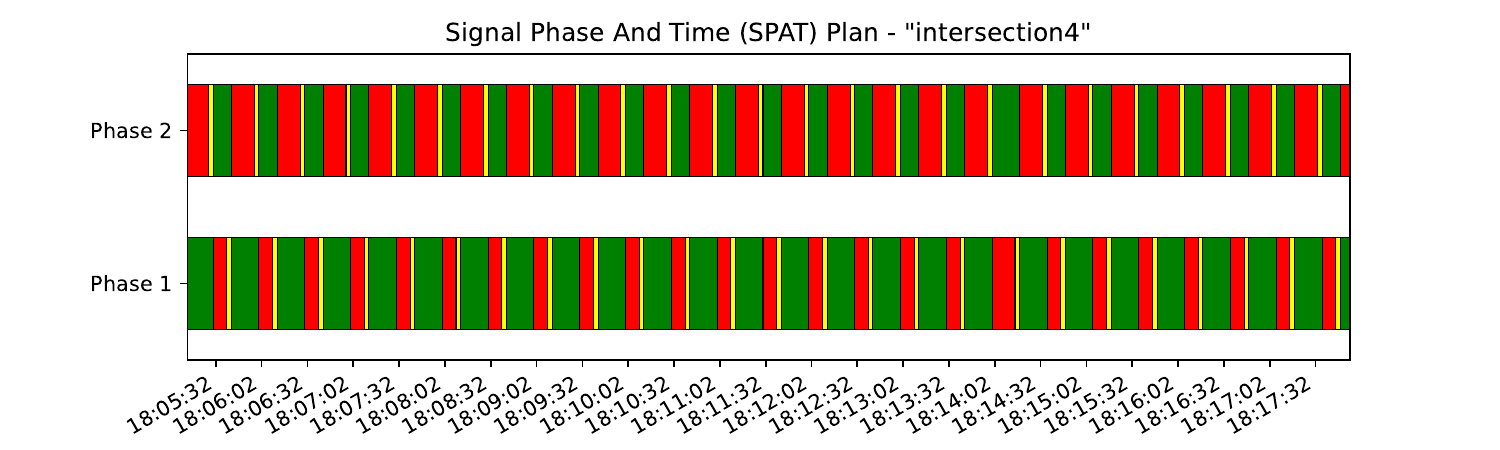}
    \includegraphics[width=0.7\linewidth]{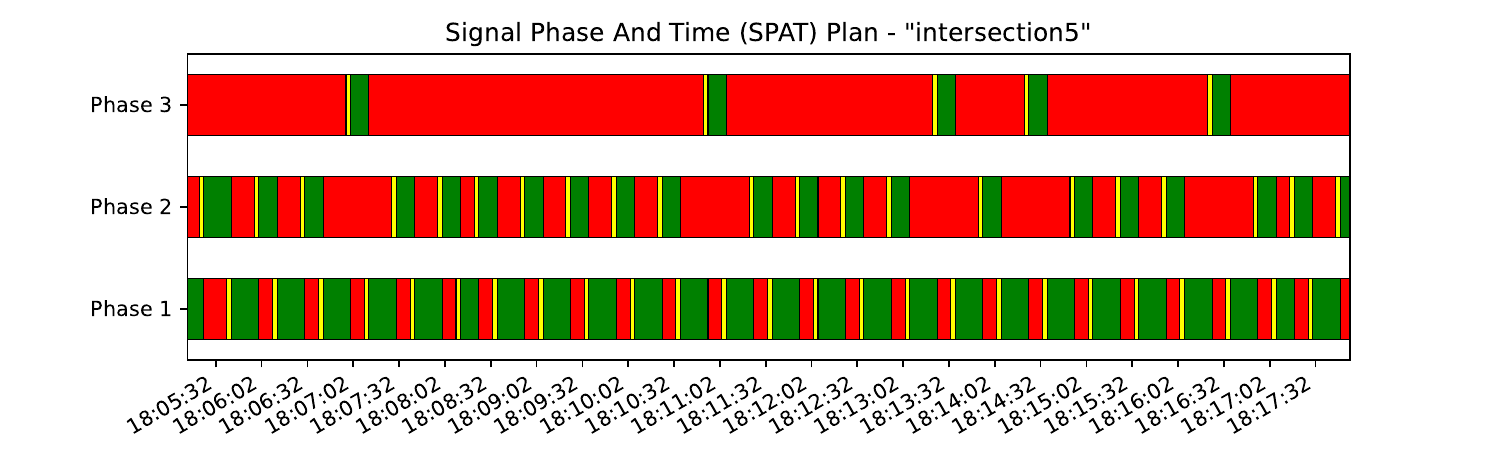}
    \caption{\textbf{Demonstration Max-Pressure (Flexible).}}
    \label{fig:demo_maxflex}
\end{figure}

Figure~\ref{fig:demo_maxflex} shows the signal phase and time (SPAT) plans for all four intersections (zoomed in for 3.5 minutes).
Contrary to the previous (fixed) version, the order of phases here is flexible, as well as the duration.

{\small
\begin{verbatim}
    intersection1 = Intersection(
        tl_id="intersection1",
        phases=[0, 2, 4],
        links={
            0: ["921020465#1_3","921020465#1_2","921020464#0_1","921020464#1_1",
                "38361907_3","38361907_2","-1164287131#1_3","-1164287131#1_2", ],
            2: ["-1169441386_2","-1169441386_1","-331752492#1_2","-331752492#1_1",
                "-331752492#0_1","-331752492#0_2",],
            4: ["-183419042#1_1","26249185#30_1","26249185#30_2","26249185#1_1",
                "26249185#1_2",],
        }
    )
    intersection2 = Intersection(
        tl_id="intersection2",
        phases=[0, 2, 4],
        # links = {0:["183049933#0_1", "-38361908#1_1"],
        #           2:["-38361908#1_1", "-38361908#1_2"],
        #           4:["-25973410#1_1", "758088375#0_1", "758088375#0_2"]},
        sensors={
            0: ["e2_183049933#0_1", "e2_-38361908#1_1"],
            2: ["e2_-38361908#1_1", "e2_-38361908#1_2"],
            4: ["e2_-25973410#1", "e2_758088375#0_1", "e2_758088375#0_2"],
        }
    )
    intersection3 = Intersection(
        tl_id="intersection3",
        phases=[0, 2, 4],
        links={
            0: ["E3_1", "-758088377#1_1", "-758088377#1_2", "-E1_1", "-E1_2"],
            2: ["E3_1", "E3_2"],
            4: ["-758088377#1_1", "-E1_1", "-E4_1", "-E4_2"],
        }
    )
    intersection4 = Intersection(
        tl_id="intersection4",
        phases=[0, 2],
        links={
            0: ["22889927#0_1", "758088377#2_1", "-22889927#2_1"], 
            2: ["-25576697#0_0"]
        }
    )
    intersection5 = Intersection(
        tl_id="intersection5",
        phases=[0, 2, 4],
        links={
            0: ["E6_1", "E6_2", "E5_1", "130569446_1", "E15_1", "E15_2"],
            2: ["E15_2", "E6_3", "E5_2", "130569446_2"],
            4: ["E10_1", "E9_1", "1162834479#1_1", "-208691154#0_1", "-208691154#1_1"],
        }
    )
    
    max_pressure_params = {
        "T_A": 5,
        "T_L": 3,  # Yellow Time
        "G_T_MIN": 5,  # Min Greentime (used for Max. Pressure)
        "G_T_MAX": 50,  # Max Greentime (used for Max. Pressure)
        "measurement_period": int(1 / 0.25),  # int(1 / simulation.time_step)
    }
    controller1 = MaxPressure_Fix(params=max_pressure_params, intersection=intersection1)
    controller2 = MaxPressure_Fix(params=max_pressure_params, intersection=intersection2)
    controller3 = MaxPressure_Fix(params=max_pressure_params, intersection=intersection3)
    controller4 = MaxPressure_Fix(params=max_pressure_params, intersection=intersection4)
    controller5 = MaxPressure_Fix(params=max_pressure_params, intersection=intersection5)
\end{verbatim}
}

\newpage
\subsubsection{Scoot/Scats}

\begin{figure} [!h]
    \centering
    \includegraphics[width=1.0\linewidth]{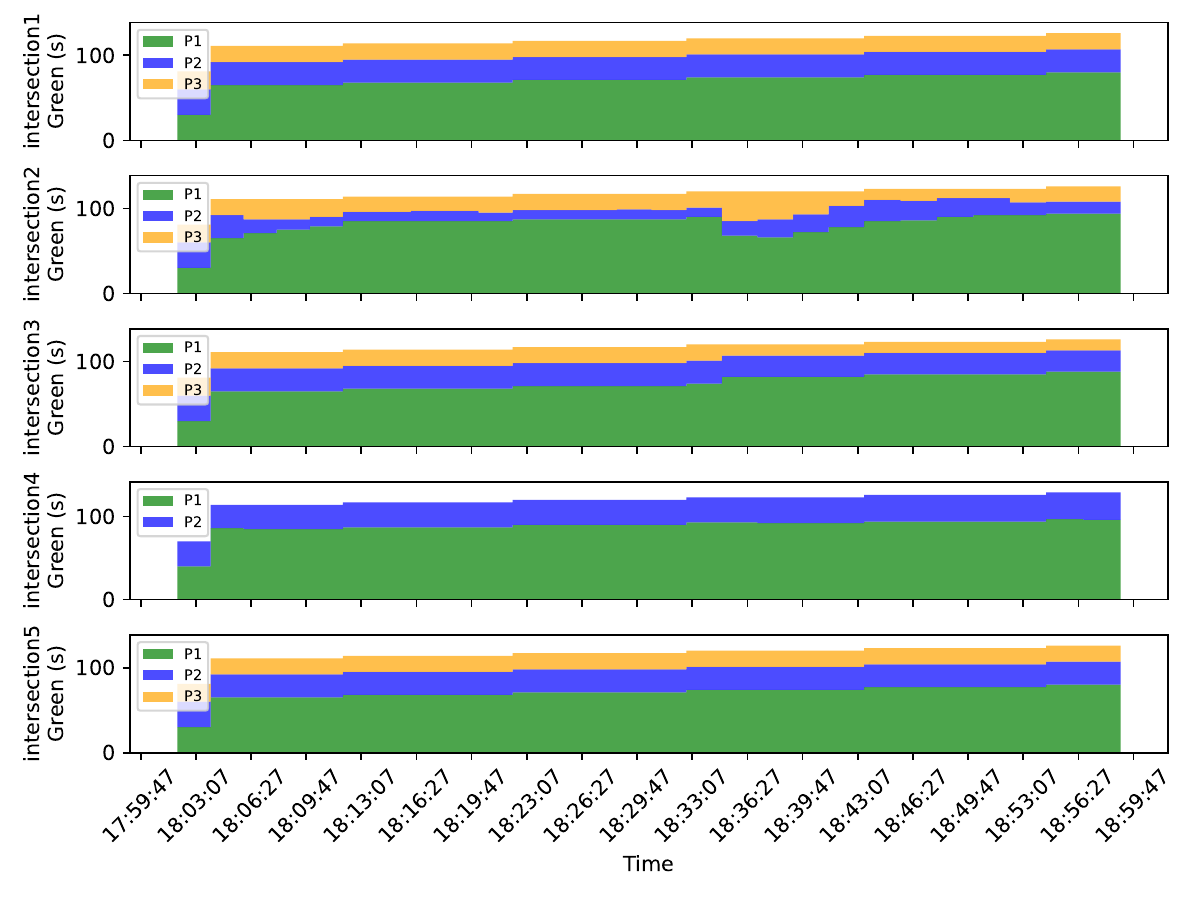}
    \caption{\textbf{Demonstration Scoot/Scats.}}
    \label{fig:demo_scosca}
\end{figure}

Figure~\ref{fig:demo_scosca} shows the signal schedules over time.
Contrary to the previous fixed version of \texttt{Max-Pressure}, the cycle length can change over time (not only green splits) for \texttt{Scoot/Scats}.
The offset optimiser was not invoked as the congestion gap did not exceed the relevant threshold.

{\small
\begin{verbatim}
    intersection1 = Intersection(
        tl_id="intersection1",
        phases=[0, 2, 4],
        links={
            0: ["921020465#1_3","921020465#1_2","921020464#0_1","921020464#1_1",
                "38361907_3","38361907_2","-1164287131#1_3","-1164287131#1_2", ],
            2: ["-1169441386_2","-1169441386_1","-331752492#1_2","-331752492#1_1",
                "-331752492#0_1","-331752492#0_2",],
            4: ["-183419042#1_1","26249185#30_1","26249185#30_2","26249185#1_1",
                "26249185#1_2",],
        },
        green_states=["GGrrGrr", "rrGGrrr", "rrrrrGG"],
        yellow_states=["yyrryrr", "rryyrrr", "rrrrryy"],
    )
    intersection2 = Intersection(
        tl_id="intersection2",
        phases=[0, 2, 4],
        links = {0:["183049933#0_1", "-38361908#1_1"],
                 2:["-38361908#1_1", "-38361908#1_2"],
                 4:["-25973410#1_1", "758088375#0_1", "758088375#0_2"]},
        green_states=["GGrrrGrr", "GGGrrrrr", "rrrGGrGG"],
        yellow_states=["yyrrryrr", "yyyrrrrr", "rrryyryy"],
    )
    intersection3 = Intersection(
        tl_id="intersection3",
        phases=[0, 2, 4],
        links={
            0: ["E3_1", "-758088377#1_1", "-758088377#1_2", "-E1_1", "-E1_2"],
            2: ["E3_1", "E3_2"],
            4: ["-758088377#1_1", "-E1_1", "-E4_1", "-E4_2"],
        },
        green_states=["GGGrrrr", "rrGGrrr", "GrrrGGG"],
        yellow_states=["yyyrrrr", "rryyrrr", "yrrryyy"],
    )
    intersection4 = Intersection(
        tl_id="intersection4",
        phases=[0, 2],
        links={
            0: ["22889927#0_1", "758088377#2_1", "-22889927#2_1"], 
            2: ["-25576697#0_0"]
        },
        green_states=["GgrrGG", "rrGGGr"],
        yellow_states=["yyrryy", "rryyyr"],
    )
    intersection5 = Intersection(
        tl_id="intersection5",
        phases=[0, 2, 4],
        links={
            0: ["E6_1", "E6_2", "E5_1", "130569446_1", "E15_1", "E15_2"],
            2: ["E15_2", "E6_3", "E5_2", "130569446_2"],
            4: ["E10_1", "E9_1", "1162834479#1_1", "-208691154#0_1", "-208691154#1_1"],
        },
        green_states=["GGrrrrGGrrrr", "rrGrrrrrGrrr", "rrrGGgrrrGGg"],
        yellow_states=["yyrrrryyrrrr", "rryrrrrryrrr", "rrryyyrrryyy"],
    )
    districts = {
        "front": ["intersection1", "intersection2"],
        "middle": ["intersection3", "intersection4"],
        "back": ["intersection5"],
    }
    critical_district_order = {
        "front": [
            "intersection1",
            "intersection2",
            "intersection3",
            "intersection4",
            "intersection5",
        ],
        "middle": [
            "intersection3",
            "intersection2",
            "intersection4",
            "intersection1",
            "intersection5",
        ],
        "back": [
            "intersection5",
            "intersection4",
            "intersection3",
            "intersection2",
            "intersection1",
        ],
    }
    connection_between_intersections = {
        "intersection1": ["183049934_1", "183049933#0_1", "1164287131#0_1"],  # To Int 2
        "intersection2": ["38361908#1_1", "E3_1"],  # To Int 3
        "intersection3": [
            "E1_1",
            "758088377#1_1",
            "758088377#2_1",
            "22889927#0_1",
        ],  # To Int 4
        "intersection4": [
            "22889927#2_1",
            "22889927#3_1",
            "22889927#4_1",
            "387296014#0_1",
            "387296014#1_1",
            "696225646#1_1",
            "696225646#2_1",
            "696225646#3_1",
            "130569446_1",
            "E5_1",
            "E6_1",
        ],  # To Int 5
    }
    intersection_group = IntersectionGroup(
        intersections=[
            intersection1,
            intersection2,
            intersection3,
            intersection4,
            intersection5,
        ],
        districts=districts,
        critical_district_order=critical_district_order,
        connection_between_intersections=connection_between_intersections,
    )
    initial_greentimes = {
        "intersection1": [30, 30, 21],
        "intersection2": [30, 30, 21],
        "intersection3": [30, 30, 21],
        "intersection4": [40, 30],
        "intersection5": [30, 30, 21],
    }
    scosca_params = {
        "adaptation_cycle": 30,
        "adaptation_green": 10,
        "green_thresh": 2,
        "adaptation_offset": 1,
        "offset_thresh": 0.5,
        "min_cycle_length": 50,
        "max_cycle_length": 180,
        "ds_upper_val": 0.925,
        "ds_lower_val": 0.875,
        "measurement_period": int(1 / 0.25),  # 1 / simulation_step_size
        "travel_time_adjustments": {
            "intersection1": ["183049934_1", 2],
            "intersection3": ["E1_1", 3],
            "intersection4": ["22889927#3_1", 9],
        },
        "intersection_offset_rules": {
            "intersection2": {
                "base_offset_from": None,
                "travel_time_from": "intersection2",
            },
            "intersection4": {
                "base_offset_from": None,
                "travel_time_from": "intersection3",
            },
            "intersection1": {
                "base_offset_from": "intersection2",
                "travel_time_from": "intersection1",
            },
            "intersection3": {
                "base_offset_from": "intersection4",
                "travel_time_from": "intersection4",
            },
            "default": {
                "base_offset_from": "intersection4",
                "travel_time_from": "intersection4",
            },
        },
    }
    controller = ScootScats(
        scosca_params, intersection_group, initial_greentimes, initial_cycle_length=120
    )
\end{verbatim}
}

\end{document}